\documentclass[]{jfm}

\usepackage[utf8]{inputenc}			
\usepackage[T1]{fontenc}			
\usepackage[british]{babel}	        
\usepackage{amsmath}				
\usepackage{amssymb}				
\usepackage{mathtools}				
\usepackage{bm}						
\usepackage{graphicx}				
\usepackage{xcolor} 				
\usepackage{comment} 				
\usepackage{array}					
\usepackage{tabularx}				
\usepackage[english=british]{csquotes} 
\usepackage{gensymb}        
\usepackage{soul}           

\usepackage{hyperref}		
\hypersetup{hidelinks} 		
\hypersetup{
    colorlinks = true,
    urlcolor   = blue,
    citecolor  = black,
}

\usepackage{ragged2e}
\usepackage{newtxtext}
\usepackage{newtxmath}
\usepackage{natbib}
\newcommand{\RomanNumeralCaps}[1]
\linenumbers

\newcommand{\rev}[1]{\textcolor{black}{#1}}

\renewcommand{\Pr}		{Pr} 

\newcommand{\Ra}   		{Ra}

\newcommand{\Racrit} 	{Ra_{c}}
\newcommand{\kcrit} 	{k_{c}}
\newcommand{\lambdacrit} 	{\lambda_{c}}

\newcommand{\Nu}   		{Nu}
\renewcommand{\Re}		{Re} 

\newcommand{\dT}   		{\Delta T}
\newcommand{\dTN}   	{\Delta T_N}

\newcommand{\Tc}   	    {T_c} 
\newcommand{\Tt}   	    {T_t} 
\newcommand{\Tb}   	    {T_b} 
\newcommand{\Th}   	    {T_h} 

\newcommand{\Gs}   	    {\varGamma_s} 
\newcommand{\Gst}   	{\varGamma_{st}} 
\newcommand{\Gsb}   	{\varGamma_{sb}} 

\newcommand{\lambdaratio}   	    {\frac{\lambda_s}{\lambda_f}} 
\newcommand{\lambdaratioIL}   	    {\lambda_s \! / \! \lambda_f} 
\newcommand{\lambdaratioInv}   	    {\frac{\lambda_f}{\lambda_s}} 
\newcommand{\lambdaratioInvIL}   	    {\lambda_f \! / \! \lambda_s} 
\newcommand{\kapparatio}   	        {\frac{\kappa_s}{\kappa_f}} 
\newcommand{\kapparatioIL}   	        {\kappa_s \! / \! \kappa_f} 
\newcommand{\kapparatioInv}   	        {\frac{\kappa_f}{\kappa_s}} 

\newcommand{\tauf}   	{\tau_f} 

\renewcommand{\Gamma}   {\varGamma} 
\renewcommand{\Phi}     {\varPhi}   
\renewcommand{\Theta}   {\varTheta} 
\newcommand{\intLS}     {\varLambda_T}
\newcommand{\Thetarms}  {\varTheta_{rms}}
\newcommand{\maxDhT}    {\max \left( \Delta_h T \right)}
\newcommand{\stdT}      {\mathrm{std} \left( T \right)}
\newcommand{\Nug}       {Nu_{CHT}}

\begin{document}

\title{Effects of conjugate heat transfer on large-scale flow structures in convection}

\author{
Matti Ettel\aff{1}, 
Philipp P. Vieweg\aff{2}
  \corresp{\email{ppv24@cam.ac.uk}} and
Jörg Schumacher\aff{1, 3}
}
\affiliation{
\aff{1}Institute of Thermodynamics and Fluid Mechanics, Technische Universit\"at Ilmenau, Postfach 100565, D-98684 Ilmenau, Germany.
\aff{2}Department of Applied Mathematics and Theoretical Physics, Wilberforce Road, Cambridge, CB3 0WA, United Kingdom. 
\aff{3}Tandon School of Engineering, New York University, New York, NY 11021, USA.}

\date{\today}

\maketitle

\begin{abstract}
The constant temperature and constant heat flux thermal boundary conditions, both developing distinct flow patterns, represent limiting cases of ideally conducting and insulating plates in Rayleigh-Bénard convection (RBC) flows, respectively. This study bridges the gap in between, using a conjugate heat transfer (CHT) set-up and studying finite thermal diffusivity ratios $\kapparatioIL$ to better represent real-life conditions in experiments. A three-dimensional RBC configuration including two fluid-confining plates is studied via direct numerical simulations given a Prandtl number $\Pr=1$. The fluid layer of height $H$ and horizontal extension $L$ obeys no-slip boundary conditions at the two solid-fluid interfaces and an aspect ratio of $\Gamma=L/H=30$ while the relative thickness of each plate is $\Gs=H_s/H=15$. The entire domain is laterally periodic. Here, different $\kapparatioIL$ are investigated for moderate Rayleigh numbers $\Ra=\left\{ 10^4, 10^5 \right\}$. We observe a gradual shift of the size of the characteristic flow patterns and their induced heat and mass transfer as $\kapparatioIL$ is varied, suggesting a relation between the recently studied turbulent superstructures and supergranules for constant temperature and constant heat flux boundary conditions, respectively. Performing a linear stability analysis for this CHT configuration confirms these observations theoretically while extending previous studies by investigating the impact of a varying solid plate thickness $\Gs$. Moreover, we study the impact of $\kapparatioIL$ on both the thermal and viscous boundary layers. Given the prevalence of finite $\kapparatioIL$ in nature, this work is a starting point to extend our understanding of pattern formation in geo- and astrophysical convection flows.
\end{abstract}

\clearpage

\section{Introduction}
\label{sec:Introduction}

Thermal convection -- the buoyancy-driven transport of mass and heat -- is an omnipresent fluid flow process in nature, occurring not just in geophysical systems like clusters of clouds over the ocean \citep{Mapes1993} or the Earth's mantle \citep{Chilla2012}, but also on other planets such as the storms on Jupiter \citep{Young2017} and Saturn \citep{Garcia2013} or stars like in the Sun's solar convection zone \citep{Schumacher2020}. Understanding this process is thus vital to comprehending geo- and astrophysical flows. 

Rayleigh-Bénard convection can be seen as the paradigm of such, containing all essential ingredients and permitting the investigation of even complex convection phenomena like pattern formation in detail. There, fluid is confined between two parallel, horizontally-extended plates while being heated from below and cooled from above. When interacting with gravity, an interplay between buoyant and viscous forces occurs which is quantified by the Rayleigh number $\Ra$. Once thermal driving gets strong enough to destabilise the fluid layer -- marked by passing the critical Rayleigh number $\Racrit$ -- instabilities start growing and convection sets in \citep{Rayleigh1916}. 

By virtue of rapid improvements in computational power, it only became possible in recent years to numerically study large-scale pattern formation for extended domains due to the strong scale separation towards the small Kolmogorov or Batchelor scales \citep{Scheel2013}. Extended fluid domains, i.e., domains possessing a (horizontal) aspect ratio $\Gamma=L/H \gg 1$ where $L$ and $H$ are the horizontal and vertical extent, respectively, are vital for understanding natural systems. Since the influence of lateral boundaries decreases with $\textit{O} \left(\Gamma^{-2}\right)$ \citep{Manneville2006, Cross2009, Koschmieder1993}, it is typically assumed that $\Gamma \gtrsim 20$ lets this impact practically vanish and thus approximates real-life scenarios fairly well \citep{Stevens2018, Koschmieder1993, Krug2020}.

Traditionally, the heating and cooling of the fluid has been achieved in two ways: either applying (two different) constant temperatures (i.e., Dirichlet type boundary condition) or a constant heat flux (i.e., Neumann type). Previous studies have found that these thermal boundary conditions govern the pattern formation process, leading to either \textit{turbulent superstructures} -- an arrangement of large-scale convection rolls and cells of size $\varLambda \sim \textit{O} \left( H \right)$ -- in the Dirichlet case \citep{Pandey2018, Stevens2018} or a pair of \textit{supergranules} -- two pairs of larger-scale convection rolls $\varLambda \sim \Gamma H \gg \textit{O} \left( H \right)$ that span eventually across the entire domain -- in the Neumann case \citep{Vieweg2021, Vieweg2022, Vieweg2024, Vieweg2024a}. Interestingly, both kinds of these turbulent long-living large-scale flow structures \citep{Vieweg2023a} are reminiscent of the respective critical pattern present at the onset of convection \citep{Rayleigh1916, Pellew1940, Hurle1967}.

\begin{figure}
\centering
\includegraphics[scale = 1.0]{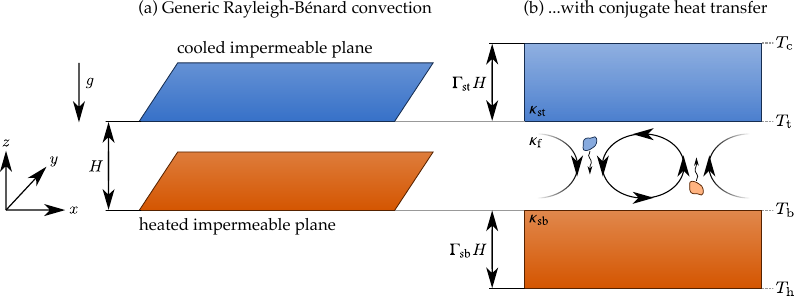}
\caption{\justifying{Fundamental configuration.
In Rayleigh-Bénard convection, (a) a layer of fluid is confined between a heated bottom and a cooled top plane, respectively. While these planes are typically also the limits of the numerical domain, (b) this study includes the (otherwise omitted) adjacent plates together with the coupled or CHT across the two solid-fluid interfaces.
The location of different temperatures is defined on the right while only $\Th$ and $\Tc$ are controlled -- other temperatures manifest dynamically. 
In this study, $\kappa_{st} = \kappa_{sb} = \kappa_{s}$ and $\Gamma_{st} = \rev{\Gamma_{sb}} = \Gs$.
}}
\label{fig:schematic_configuration}
\end{figure}

These two conditions, however, represent mathematically idealised scenarios that are illustrated in figure \ref{fig:schematic_configuration} (a). In fact, they omit the vertically adjacent matter or solid that confines and, thus, influences the fluid. Using a \textit{conjugate heat transfer} (CHT) set-up addresses the problem more holistically by modelling these plates as solid thermal conductors such that not only the heat transfer in the fluid, but also in the adjacent solids as well as their solid-fluid interaction is considered \citep{Perelman1961}. Here, as illustrated in figure \ref{fig:schematic_configuration} (b), (external) thermal boundary conditions are applied as constant temperatures $\Th$ and $\Tc$ at the very bottom and top of the solid plates of relative thickness or \textit{vertical aspect ratio} $\Gs=H_s/H$, respectively. This causes a temperature gradient within the solid plates before reaching the (internal) thermal boundary conditions at the solid-fluid interfaces -- offering the, based on system dynamics, dynamically manifesting temperatures $\Tb$ and $\Tt$ -- where both the temperature and heat flux are coupled between the different sub-domains. 

The ratio of thermal diffusivities between the solids and the fluid, $\kapparatioIL$, plays an integral role in governing the aforementioned pattern formation process as the Neumann case is represented by $\kapparatioIL \rightarrow 0$ whereas the Dirichlet case corresponds to $\kapparatioIL \rightarrow \infty$. While previous studies in large aspect ratios \citep{Pandey2018, Schneide2022, Vieweg2021, Vieweg2021a, Vieweg2022, Vieweg2024, Vieweg2025, Vieweg2023a, Vieweg2024a} have only focused on these extreme ends, natural systems always are located somewhere in between. Although some studies of finite $\kapparatioIL$ have already analysed heat transfer in a CHT set-up experimentally \citep{Vasilev2015} or for cylindrical cells of small aspect ratios $\Gamma=1/2$ \citep{Verzicco2002, Verzicco2004, Foroozani2021} \rev{and others have contrasted the difference in heat transport between the Dirichlet and Neumann cases in small cells} \rev{\citep{Verzicco2008, Johnston2009}}, the pattern formation process has not yet been investigated. Since both turbulent superstructures \citep{Krug2020} and supergranules \citep{Vieweg2021} are of crucial importance for the induced heat transfer across the fluid layer, a detailed understanding of pattern formation is indispensable. 
\rev{Moreover, the ratio of thermophysical properties between the solid and fluid is crucial beyond our focus on pattern formation aspects, e.g., for the turbulent heat transfer across the fluid layer. This holds particularly for laboratory experiments at very large Rayleigh numbers of $\Ra\gtrsim 10^{12}$ where the enhanced turbulence-induced effective conductivity in the fluid can be close to the one in the plates. This requires corrections in $Nu$, which have been discussed for example by \citet{Niemela2006}.}

This study aims to represent natural scenarios of $\kapparatioIL \in \left( 0, \infty \right)$ at a large aspect ratio $\Gamma=30$ more accurately by considering the coupled or conjugate heat transfer at the solid-fluid interfaces, thus coining the term \textit{natural thermal boundary conditions}. 
Comprehending this region is crucial to enhancing our understanding of convection flows and their properties in real-life geo- and astrophysical scenarios. To do so, direct numerical simulations are conducted for two Rayleigh numbers $\Ra=\left\{10^4,10^5\right\}$ over an array of different $\kapparatioIL$. As a primary result of this work, we observe pronounced gradual shifts of both the size of flow structures and their induced heat transfer when varying $\kapparatioIL$, underlining the importance of \textit{long-living large-scale flow structures} as an umbrella term for both turbulent superstructures and supergranules. The transition of large-scale flow structures comes with gradual shifts in both the thermal as well as viscous boundary layer thicknesses.
This numerical approach is complemented theoretically by a comprehensive \textit{linear stability analysis} regarding the onset of convection. This analysis confirms the aforementioned transition between flow structures as a gradual shift towards larger critical wavenumbers $\kcrit$ is observed when moving from Neumann to Dirichlet conditions.
As the vertical aspect ratio or plate thickness $\Gs$ represents an additional free parameter in the CHT set-up, we extend both of our approaches towards a variation of $\Gs$ and find that thin plates stabilise the layer especially for moderate $\lambdaratioIL$. We remark that this, \rev{together with easy-to-handle regression fits,} extends the results obtained by \citet{Hurle1967} for infinitely thick plates. 
This study bridges the gap between classical thermal boundary conditions by incorporating solid sub-domains together with the coupled temperatures and heat transfers at the solid-fluid interfaces, allowing to interpret natural flows in the geo- and astrophysical context more successfully.

\section{Governing equations and numerical method}
\label{sec:Governing_equation_and_numerical_method}

\subsection{Governing equations}
\label{subsec:Governing_equations}

We consider an incompressible flow based on the Oberbeck-Boussinesq approximation \citep{Oberbeck1879, Boussinesq1903}. This means that all material parameters are constant -- except for the mass density, the latter of which varies at first order with temperature when interacting with gravity only. 
The three-dimensional equations of motion are non-dimensionalised based on the fluid layer height $H$ and the temperatures at the bottom and top of this fluid layer, $\Tb$ and $\Tt$ (see also figure \ref{fig:schematic_configuration} (b)), respectively. We use the spatio-temporal average of these temperature fields to define the characteristic (dimensional) temperature scale $\dT := \langle \Tb - \Tt \rangle_{A,t}$ where $A$ denotes the entire horizontal cross-section. Note that this implies a non-dimensionalisation $T = \dT \thickspace \tilde{T}$ together with the assumption of a resulting non-dimensional temperature difference across the fluid layer of $\dTN := \langle \tilde{\Tb} - \tilde{\Tt} \rangle_{A,t} \equiv 1$. \rev{We stress explicitly that we write out the tildes here to clearly distinguish the dimensional $\dT$ and non-dimensional temperature difference $\dTN$ but will, from now on, mostly omit such for better readability.}
By virtue of the free-fall inertia balance, the free-fall velocity $U_f = \sqrt{\alpha g \dT H}$ and time scale $\tauf = H / U_f = \sqrt{H / \alpha g \dT}$ can be acquired where $\alpha$ is the volumetric thermal expansion coefficient of the fluid at constant pressure, $g$ the acceleration due to gravity, and $\rho_{ref,f}$ the reference density of the fluid at reference temperature. 
We are going to solve the resulting coupled equations using the spectral-element solver Nek5000 \citep{Fischer1997,Scheel2013}.

For the fluid sub-domain, the governing equations are 
\begin{align}
\label{eq:CE}
\nabla \cdot \bm{u} &= 0 , \\
\label{eq:NSE}
\frac{\partial \bm{u}}{\partial t} + \left( \bm{u} \cdot \nabla \right) \bm{u} &= - \nabla p + \sqrt{\frac{\Pr}{\Ra}}  \nabla^{2} \bm{u} + T \bm{e}_{z} , \\
\label{eq:EE_fl}
\frac{\partial T}{\partial t} + \left( \bm{u} \cdot \nabla \right) T &= \frac{1}{\sqrt{\Ra \Pr}}  \nabla^{2} T.
\end{align}
In contrast, the solid sub-domains require to solve a pure diffusion equation 
\begin{align}
\label{eq:EE_s}
\frac{\partial T}{\partial t} &= \kapparatio \frac{1}{\sqrt{\Ra \Pr}}  \nabla^{2} T
\end{align}
only \citep{Foroozani2021, Vieweg2025}.
Here, $\bm{u}$, $T$ and $p$ represent the (non-dimensional) velocity, temperature and pressure field, whereas $\kappa_{\Phi} = \lambda_{\Phi} / \rho_{{ref},\Phi} c_{p,\Phi}$ is the thermal diffusivity. Its ratio between the solid and fluid domains $\kapparatioIL$ constitutes an important control parameter over the course of this work, with the subscripts $\Phi=\{f,s\}$ denoting the fluid and solid, respectively. $\lambda_{\Phi}$ represents the thermal conductivity, $\rho_{\Phi}$ the mass density, and $c_{p,\Phi}$ the specific heat capacity at constant pressure. Furthermore, the Rayleigh and Prandtl number are defined via
\begin{equation}
\label{eq:def_Rayleigh_number_Prandtl_number}
\Ra = \frac{\alpha g \dT H^{3}}{\nu_f \kappa_f} \qquad \textrm{and} \qquad \Pr = \frac{\nu_f}{\kappa_f},
\end{equation}
where $\nu_f$ is the kinematic viscosity of the fluid .
Note that equation \eqref{eq:EE_s} holds for both the top and bottom plate -- as they will offer identical thermal diffusivities -- and, \rev{thus, differs} from our recent work \citep{Vieweg2025}.

\subsection{Numerical domain, boundary and initial conditions}
\label{subsec:Numerical_domain_boundary_and_initial_conditions}

\begin{table}
\begin{center}
\def~{\hphantom{0}}
  \begin{tabular}{llrr}
       Setting & Solid / Fluid & $\kapparatioIL$ & $\lambdaratioIL$    \\ [3pt]
            Ocean & \rev{Seawater} / Air & $5.71 \times 10^{-3}$ & 22.8 \\
            \rev{Mountain (Dolomites)} & \rev{Dolomite / Air} & \rev{$1.16 \times 10^{-1}$} & \rev{220} \thickspace \thickspace\\
            Beach & Quartz / Air & $2.20 \times 10^{-1}$ & 352 \thickspace \thickspace
  \end{tabular}
  \caption{\justifying{Ratios of thermophysical properties in natural configurations. Values of both the thermal diffusivity as well as thermal conductivity are taken at $10\degree \mathrm{C}$ \rev{for seawater (salinity of 35 ppt), air and quartz from \citet{Ochsner2019, Ibrahim2014} and for dolomite from \citet{Horai1971, Stout1963}}.
  }}
  \label{tab:natural_kapparatio_lambdaratio}
  \end{center}
\end{table}

These governing equations are complemented by a numerical domain and its respective boundary conditions. We define the horizontal extent $L$ of our numerical domain by the (horizontal) aspect ratio $\Gamma:=L/H$, whereas the vertical aspect ratio $\Gs:=H_s/H$ describes the thickness of each of the surrounding solid plates. Note that both of these aspect ratios are based on the height $H$ of the fluid domain, whereas any sub-domain offers the square horizontal cross-section $A=\Gamma \times \Gamma$. Regarding figure \ref{fig:schematic_configuration}, the solid bottom and top domains are thus situated at $z \in [-\Gs,0]$ and $z \in [1,1+\Gs]$, respectively, with the fluid in between at $z \in [0,1]$.

We consider a horizontally periodic domain where any quantity $\bm{\Phi}$ repeats according to
\begin{equation}
    \label{eq:BC_periodic}
   \bm{\Phi} \left( \bm{x} \right) = \bm{\Phi} \left( \bm{x} + i_x L_x \bm{e}_{x} + i_y L_y \bm{e}_{y} \right), \quad i_{x,y} \in \mathbb{N}
\end{equation}
and which offers no-slip boundary conditions
\begin{equation}
    \label{eq:BC_no_slip}
    \bm{u} \left(z=\{0,1\} \right) = \rev{\bm{0}}.
\end{equation}
at both solid-fluid interfaces. Thermal boundary conditions are applied in the form of constant temperatures at the very top and bottom of the plates (i.e., $z=\{-\Gs, 1+\Gs\}$) which will further be referred to as
\begin{alignat}{2}
    \label{eq:Th}
    &T \left( z = - \Gs \right) &&= \Th \quad \textrm{and}\\
    \label{eq:Tc}
    &T \left( z = 1 + \Gs \right) &&= \Tc.
\end{alignat}

By nature of the CHT set-up, temperature fields and diffusive heat fluxes are coupled at the solid-fluid interfaces (i.e., $z=\left\{ 0,1 \right\}$) according to
\begin{align}
    \label{eq:tBC_z0}
    \Tb := T_f \left( z=0 \right) = T_{s} \left( z=0 \right), & \quad \lambdaratio \left. \frac{\partial T_{s}}{\partial z}\right\rvert_{z=0} = \left. \frac{\partial T_{f}} {\partial z}\right\rvert_{z=0}, \\
    \label{eq:tBC_z1}
    \Tt := T_f \left( z=1 \right) = T_{s} \left( z=1 \right), & \quad \lambdaratio \left. \frac{\partial T_{s}}{\partial z}\right\rvert_{z=1} = \left. \frac{\partial T_{f}} {\partial z}\right\rvert_{z=1}.
\end{align}
Note that while the energy equation \eqref{eq:EE_s} contains $\kapparatioIL$ due to the non-dimensionalisation, the boundary conditions \eqref{eq:tBC_z0} and \eqref{eq:tBC_z1} include the ratio $\lambdaratioIL$ to match the diffusive heat fluxes at the interfaces. 
\rev{In order to avoid another control parameter, we assume in our simulations $\rho_s c_{p,s} / \rho_f c_{p,f} = 1$ such that $\lambdaratioIL \equiv \kapparatioIL$ follows.}
We will thus use $\kapparatioIL$ as the control parameter for the discussions in the main text, except for the linear stability analysis. 
We additionally stress that while we fix the externally applied temperatures $\Th$ and $\Tc$, see again eqns. \eqref{eq:Th} and \eqref{eq:Tc}, the resulting temperature fields at the solid-fluid interfaces $\Tb$ and $\Tt$ vary in both space and time by virtue of the systems dynamics. 

The ratio $\kapparatioIL$ strongly impacts the way in which the fluid interacts with the solid and vice versa. For $\kapparatioIL \rightarrow \infty$, the Dirichlet case is resembled where the temperatures at the solid-fluid interfaces become constant (since the solid is a much better thermal conductor). In contrast, for $\kapparatioIL \rightarrow 0$ the Neumann case is mimicked where the vertical temperature gradient becomes constant at these interfaces (since thermal resistance through the fluid is smaller compared to the solid). 
A more elaborate explanation is provided in appendix \ref{sec:Appendix_convergence}.
In this study, bridging the gap between Dirichlet and Neumann conditions, we are interested in a broad range of $\kapparatioIL$ centred around unity. Table \ref{tab:natural_kapparatio_lambdaratio} includes this ratio for a variety of natural configurations and shows that natural flows tend to offer $\kapparatioIL \sim \textit{O} \left(10^{-3}...10^{-1} \right)$.

Concerning our initial condition, we follow the procedure described and introduced by \cite{Vieweg2025}. In a nutshell, we initialise each simulation with a fluid is at rest, i.e., $\bm{u} \left( \bm{x}, t=0 \right) = 0$, and linear temperature profiles which respect the internal boundary conditions between the different sub-domains as outlined in equations \eqref{eq:tBC_z0} and \eqref{eq:tBC_z1}. By adding some tiny random thermal noise $0 \leq \varUpsilon \leq 10^{-3}$, we accelerate the transition to the statistically stationary state under assumption of an initial Nusselt number $\Nu \left(t=0 \right) > 1$ based on preliminary simulation runs.

\rev{The required, externally applied temperatures $\Th$ and $\Tc$ (see eqs. \eqref{eq:Th} and \eqref{eq:Tc}) are determined as follows. 
First, we define the global Nusselt number}
\begin{align}
\label{eq:def_Nusselt_number}
\nonumber
\rev{\Nu (t)
:= \frac{\left\langle \left(\bm{J}_{dif} + \bm{u} T \right) \cdot \bm{e}_{z} \right\rangle_{V_f}}{\langle \bm{J}_{dif} \cdot \bm{e}_{z} \rangle_{V_f}} }
& \rev{ = \left\langle - \frac{\partial T }{\partial z} \right\rangle_{V_f}+ \left\langle \sqrt{\Ra \Pr} ~ u_{z} T \right\rangle_{V_f} } \\
&\rev{ = 1 + \sqrt{\Ra \Pr} \left\langle u_{z} T \right\rangle_{V_f} ,}
\end{align}
\rev{as a measure of the induced amplification of the global heat transfer due to convective fluid motion. Note that the latter is associated with $\bm{u} T$ and in contrast to the diffusive heat current $\bm{J}_{dif}$ while $V_f$ represents the fluid volume \citep{Otero2002, Vieweg2023a}.}
\rev{Second, given the linear conduction profiles outlined in \citep{Vieweg2025} and an assumed amplification of heat transfer as measured by $\Nu$, it is possible to estimate these applied temperatures -- that are required to achieve $\dTN \! \approx \! 1$ -- according to}
\begin{equation}
    \label{eq:Th_Tc_conduction_profiles}
    \rev{\Th = 1 + \lambdaratioInv \Gs \Nu \quad \textrm{and} \quad
    \Tc = - \lambdaratioInv \Gs \Nu.}
\end{equation}
\rev{Note that, as the final $\Nu$ is not known a-priori, either preliminary 2- and 3-dimensional simulation runs or our introduced $\tanh$-relationship (see table \ref{tab:regression_parameters_Nu_Re}) have been used to find appropriate values for $\Th$ and $\Tc$.}

\section{Linear stability analysis of the coupled system}
\label{sec:Linear_stability_analysis_of_the_coupled_system}

Flow structures at the onset of convection, as derivable analytically via a linear stability analysis, are a characteristic of the dynamical system and, thus, helpful for understanding even turbulent flow structures \citep{Pandey2018, Vieweg2021}. 
Although such a linear stability analysis yields a relation $\Ra (k)$, it is the global minimum of this function that determines the critical Rayleigh number $\Racrit$ and critical wave number $\kcrit$. If $\Ra \gtrsim \Racrit$, linear perturbations grow exponentially over time and the associated size or wavelength of the emerging flow structures $\lambdacrit = 2 \pi / \kcrit$.

While, given our no-slip boundary conditions, ($\Racrit$, $\kcrit$) are well-known for the Dirichlet ($\Racrit=1707.8, \kcrit=3.13$) \citep{Pellew1940} and Neumann ($\Racrit=6!=720, \kcrit=0$) \citep{Hurle1967} cases, these values change for vertically infinitely extended (i.e., $\Gs \rightarrow \infty$) CHT set-ups with finite ratios of thermal conductivities $\lambdaratioIL \in \left( 0, \infty \right)$ \citep{Hurle1967}. One can expect that a finite plate thickness $\Gs$ adds an additional layer of complexity. 

This study extends the work of \citet{Hurle1967} by (i) investigating a broader range of $\lambdaratioIL$, (ii) considering even finite plate thicknesses $\Gs$ (in section \ref{sec:The_impact_of_the_plate_thickness}), and (iii) deriving easy-to-handle relationships between $\Racrit$, $\kcrit$ and $\lambdaratioIL$.

In order to determine the neutral stability curve and subsequently derive $\Racrit$ as well as $\kcrit$, a system of four equations needs to be solved where the determinant of the coefficient matrix $\bm{M_E}$ is set to zero to obtain a non-trivial solution,
\begin{alignat}{2}
    \label{eq:det_ME}
    &\det{\bm{M_E}} \stackrel{!}{=} 0 = \left\rvert \begin{matrix}
        0 & 1 & 1 & 1 \\
        0 & q_1t_1 & q_2t_2 & q_3t_3 \\
        - \lambdaratioInv \tanh{\left(k\Gs \right)} & \gamma & -\frac{\gamma}{2} \left( 1-\mathrm{i} \sqrt{3} \right)  & -\frac{\gamma}{2} \left( 1-\mathrm{i} \sqrt{3} \right) \\
        k & \gamma q_1t_1 & -\frac{\gamma}{2} \left( 1-\mathrm{i} \sqrt{3} \right) q_2t_2 & -\frac{\gamma}{2} \left( 1-\mathrm{i} \sqrt{3} \right) q_3t_3
    \end{matrix} \right\rvert .
\end{alignat}
In the following, we only regard the case of even modes as it provides the lower values for $\Racrit$. A detailed derivation -- explaining all involved variables -- is provided in appendix \ref{sec:Appendix_LSA}.

\begin{figure}
\centering
\includegraphics[width = \textwidth]{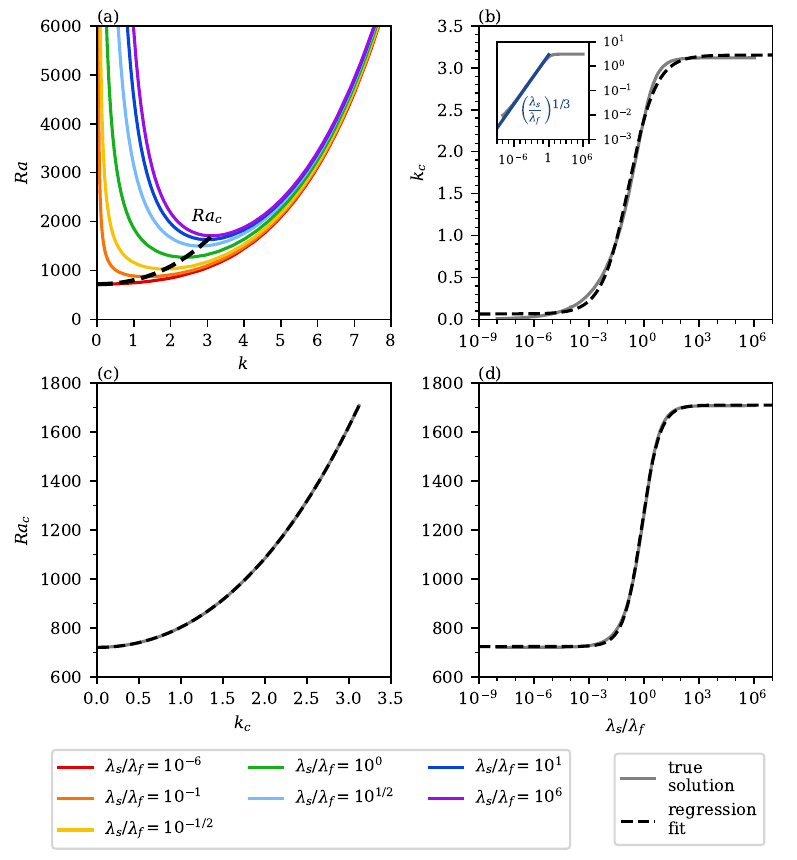}
\caption{\justifying{
Linear stability of conjugate heat transfer-driven Rayleigh-Bénard convection.
The combination of thermophysical properties controls both the general stability ($\Racrit$) as well as the size of the critical flow structures ($\kcrit$).
(a) Different neutral stability curves indicate (c) a gradual and monotonic transition of both $\Racrit$ and $\kcrit$ (see also panels (d) and (b), respectively). The true solutions from panels (b -- d) can be approximated well by tanh- or polynomial-based regressions using parameters described by table \ref{tab:regression_parameters_k_Ra}.
Note the different convergence behaviour of $\kcrit$ for $\lambdaratioIL \rightarrow \infty$ ($\kcrit = \mathrm{const.}$) and $\lambdaratioIL \rightarrow 0$ ($\kcrit \sim \left( \lambdaratioIL \right)^{1/3}$) as highlighted by the inset in panel (b), the latter of which plots the data double-logarithmically instead. 
}}
\label{fig:linear_stability}
\end{figure}

Figure \ref{fig:linear_stability} (a) contrasts different resulting neutral stability curves for different $\lambdaratioIL$ given $\Gs \rightarrow \infty$. Remember that the extreme cases $\lambdaratioIL=\left\{10^{-6}, 10^{6} \right\}$ mimic the Neumann and Dirichlet cases, respectively, and with which they correspond \citep{Pellew1940, Chandrasekhar1981, Takehiro2002}. 
Interestingly, after having solved equation \eqref{eq:det_ME} for a large number of $\lambdaratioIL$, we find a gradual and monotonic transition for both $\Racrit$ and $\kcrit$ in between these extreme conditions as visualised in panels (b--d). 
Our analysis shows that the onset of convection is generally delayed (i.e., larger $\Racrit$) with smaller emerging flow structures (i.e., larger $\kcrit$) for relatively better solid thermal conductors (i.e., increasing $\lambdaratioIL$).
As highlighted by the inset in panel (b), $\kcrit \left( \lambdaratioIL \right)$ exhibits different asymptotic convergence behaviours for the extremes of $\lambdaratioIL$: While $\kcrit \sim \lambdaratioIL^{1/3}$ for $\lambdaratioIL \rightarrow 0$, $\kcrit \simeq 3.13 = \textrm{const}$ for $\lambdaratioIL \rightarrow \infty$. Interestingly, the inflection point is around $\lambdaratioIL \approx 10^{-3/4}$ (rather than $10^{0}$) and thus introduces an asymmetry with respect to $\lambdaratioIL$.

\begin{table}
  \begin{center}
\def~{\hphantom{0}}
  \begin{tabular}{lccccccc}
       $f$                                      & $a$       & $b$   & $c$   & $d$       & $e$       & $R^2$     & \rev{range of applicability}       \\ [3pt]
       $\kcrit \left( \lambdaratioIL \right)$   & ~~1.543   & 0.293 & ~0.554 & ~~~1.608 & --        & 0.9985    & \rev{ $\lambdaratioIL \in \left[ 0, \infty \right)$ } \\
       $\Racrit \left( \lambdaratioIL \right)$  & 493.546   & 0.441 & ~0.122 & 1216.713 & --        & 0.9999    & \rev{ $\lambdaratioIL \in \left[ 0, \infty \right)$ } \\
       $\Racrit \left( \kcrit \right)$          & ~~0.341   & 7.576 & 74.868 & ~~~0.234 & 719.996   & 1.0000    & \rev{ $\kcrit         \in \left[ 0, 3.13   \right]  $ }
  \end{tabular}
  \caption{\justifying{Regression parameters for $\kcrit$ and $\Racrit$. 
  A $\tanh$-fit of the form $f \left(\lambdaratioIL, \Ra \right) = a \tanh{ \left[ b  \log{ \left( \lambdaratioIL \right) } + c \right] } + d$ is applied to the values in panels (b) and (d) of figure \ref{fig:linear_stability}. For $\Racrit \left( \kcrit \right)$ in panel (c), a fourth-order polynomial fit of the form $f \left(\kcrit \right) = a \kcrit^4 + b \kcrit^3 + c \kcrit^2 + d \kcrit + e$ is applied. $R^2$ is the coefficient of determination \citep{Wright1921} and underlines the quality of these fits.
  }
  }
  \label{tab:regression_parameters_k_Ra}
  \end{center}
\end{table}

These (true) solutions are the result of solving equation \eqref{eq:det_ME}. In order to provide handier solutions that are more accessible, we apply tanh- or polynomial-based regressions to the original data from figure \ref{fig:linear_stability} (b -- d) and include them therein. Given the parameters included in table \ref{tab:regression_parameters_k_Ra}, such simple regression fits approximate the true solutions (very) well. 

Albeit this linear stability analysis is based on $\Gs \rightarrow \infty$, we find that its solutions practically coincide with our numerically employed finite case $\Gs = 15$ as shown in section $\ref{sec:The_impact_of_the_plate_thickness}$.

\section{Non-linear pattern formation}
\label{sec:Non_linear_pattern_formation}

\subsection{Conducted simulations and pattern formation process}
\label{sec:Conducted_simulations_and_pattern_formation_process}

In order to systematically investigate the impact of natural thermal boundary conditions on convection flows \rev{beyond their onset}, we conduct two main series of simulations at two $\Ra$ varying $\kapparatioIL$ across a broad range. For all these simulations, the Prandtl number $\Pr=1$, horizontal aspect ratio $\Gamma=30$, and vertical aspect ratio $\Gs=15$. Our choice of such a horizontally extended domain makes the heat and momentum transfer, as quantified by $\Nu$ and $\Re$, independent of $\Gamma$ \citep{Stevens2018} and delays limiting the pattern formation process by the horizontal extent of the domain \citep{Stevens2018, Krug2020, Vieweg2023a} \rev{as further discussed in section \ref{sec:Quantitative_analysis_of_convective_flow_patterns}}. We report a third series of simulations at different $\Gs$ given $\kapparatioIL=10^0$ in section \ref{sec:The_impact_of_the_plate_thickness} whereas we contrast extreme cases of $\kapparatioIL$ with their plate-less classical representatives in appendix \ref{sec:Appendix_convergence}.

\begin{figure}
\centering
\includegraphics[width = \textwidth]{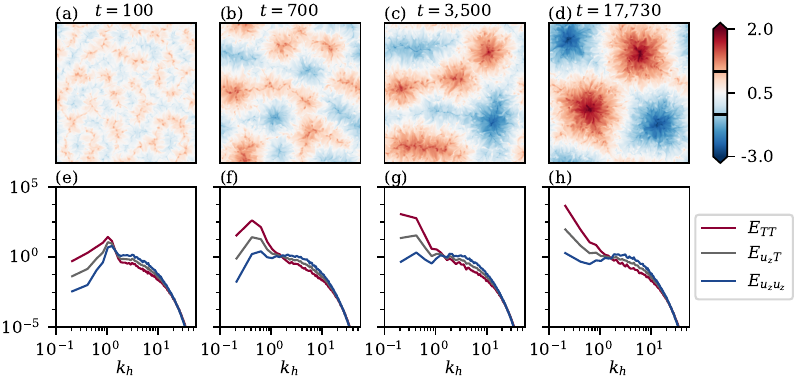}
\caption{\justifying{\rev{
Gradual pattern formation. 
(a) At early times, large-scale granulated flow structures emerge that (b, c) gradually merge and form even larger supergranules before (d) a statistically stationary state is reached. Here we visualise the thermal footprint $T \left( x, y, z = 0.5, t \right)$ of these flow structures.
(e--h) The corresponding azimuthally averaged Fourier energy spectra (of various fields) highlight a gradual shift of spectral energy towards larger horizontal scales.
This shift is governed by $\kapparatioIL$: As $\kapparatioIL \rightarrow 0$, more energy accumulates at even smaller $k$. In contrast to the idealised Neumann case -- compare with figure 3 from \citep{Vieweg2021}) -- the growth of the supergranules stops in this CHT set-up of $\Ra=10^5$ and $\kapparatioIL=10^0$ (Case $\mathrm{C5c}$) before reaching domain size.
}}}
\label{fig:Pattern_formation}
\end{figure}

Convective pattern formation is known to be a gradual process which reaches statistically stationary flow structures only after \rev{a long time, potentially even $\textit{O} \left( 10^4 \tauf \right)$ or longer} \citep{Vieweg2021, Vieweg2022, Vieweg2024, Vieweg2023a, Vieweg2024a}. 
\rev{Figure \ref{fig:Pattern_formation} illustrates this process for one of our present CHT cases: At the beginning, granulated flow structures manifest that merge over time towards larger structures. While panels (a--c) depict the transient regime, panel (d) illustrates the flow structures by means of their thermal footprint at their statistically stationary state.}
The reach of this late regime has successfully been probed using different measures such as the integral length scale \citep{Parodi2004, Vieweg2022}
\begin{equation}
\label{eq:def_integral_length_scale}
\intLS \left( z = 0.5, t \right) 
:= 2 \pi \frac{\int_{k_{h}} \left[ E_{TT} / k_{h} \right] dk_{h}}{\int_{k_{h}} E_{TT}  dk_{h}}
\end{equation}
with $E_{TT} \equiv E_{TT} \left( k_{h}, z = 0.5, t \right)$ representing the azimuthally averaged Fourier energy spectrum of the temperature field and $k_{h}$ the horizontal wavenumber, or the thermal variance \citep{Vieweg2022}
\begin{equation}
    \label{eq:Theta_rms_t}
    \Theta_{rms} \left( t \right) := \sqrt{\langle \Theta^2 \rangle_V} 
    \qquad \textrm{with} \quad 
    \Theta(\bm{x},t) := T(\bm{x},t) - T_{lin}(z)
\end{equation}
where $\Theta_{rms}$ is the temperature deviation field around the mean linear conduction profile $T_{lin} \simeq 1 - z$ across the fluid domain \citep{Vieweg2023a, Vieweg2024a}. In case of Neumann-type thermal boundary conditions, $\intLS$ usually converges more quickly than $\Thetarms$ \citep{Vieweg2023a}.

\newcommand{\hp}{\hphantom{1}}
\newcommand{\hpp}{\hphantom{.01}}
\begin{table}
\centering
\begin{tabular}{@{\hskip 0mm} l @{\hskip 6.0mm} c @{\hskip 6.0mm} l @{\hskip 6.0mm} c @{\hskip 6.0mm} c @{\hskip 6.0mm} c @{\hskip 6.0mm} r @{\hskip 6.0mm} r @{\hskip 0mm}}
        Identifier      & $\Ra$    & $\kapparatioIL$       & $\Gs$  & $N_e$                                            & $N$   & $t_r \quad$  & $\left( \Th - \Tc -1 \right) \! / 2$      \\ [3pt]
        $\mathrm{N4}$   & $10^4$    & $\rightarrow 0$       & $15$  & $100^2\!\times\!4$                                            & $7$   & $10,200$  & $\mathrm{-}$      \\
        $\mathrm{C4N}$  & $10^4$    & $10^{-6}$             & $15$  & $100^2\!\times\!\left( 4 \! + \! 2 \! \times \! 30 \right)$   & $7$   & $8,850$   & $47.5\times 10^6$  \\
        $\mathrm{C4a}$  & $10^4$    & $10^{-1}$             & $15$  & $100^2\!\times\!\left( 4 \! + \! 2 \! \times \! 30 \right)$   & $7$   & $22,950$  & $445.70$           \\
        $\mathrm{C4b}$  & $10^4$    & $10^{-1/2}$           & $15$  & $100^2\!\times\!\left( 4 \! + \! 2 \! \times \! 30 \right)$   & $7$   & $16,100$  & $128.25$          \\
        $\mathrm{C4c}$  & $10^4$    & $10^{0}$              & $15$  & $100^2\!\times\!\left( 4 \! + \! 2 \! \times \! 30 \right)$   & $7$   & $12,000$  & $36.50$            \\
        $\mathrm{C4d}$  & $10^4$    & $10^{1/4}$            & $15$  & $100^2\!\times\!\left( 4 \! + \! 2 \! \times \! 30 \right)$   & $7$   & $6,100$   & $19.90$         \\
        $\mathrm{C4e}$   & $10^4$   & $10^{1/2}$            & $15$  & $100^2\!\times\!\left( 4 \! + \! 2 \! \times \! 30 \right)$   & $7$   & $10,100$  & $10.90$            \\
        $\mathrm{C4f}$   & $10^4$   & $10^{3/4}$            & $15$  & $100^2\!\times\!\left( 4 \! + \! 2 \! \times \! 30 \right)$   & $7$   & $2,900$   & $6.14$        \\
        $\mathrm{C4g}$   & $10^4$   & $10^{1}$              & $15$  & $100^2\!\times\!\left( 4 \! + \! 2 \! \times \! 30 \right)$   & $7$   & $6,500$   & $3.45$            \\
        $\mathrm{C4D}$   & $10^4$   & $10^{6}$              & $15$  & $100^2\!\times\!\left( 4 \! + \! 2 \! \times \! 30 \right)$   & $7$   & $8,140$   & $0.00$               \\
        $\mathrm{D4}$   & $10^4$    & $\rightarrow \infty$  & $15$  & $100^2\!\times\!4$                                            & $7$   & $10,550$  & $\mathrm{-}$      \\
        \multicolumn{8}{l}{} \\
        $\mathrm{N5}$   & $10^5$    & $\rightarrow 0$       & $15$  & $100^2\!\times\!4$                                            & $11$   & $7,000$  & $\mathrm{-}$      \\
        $\mathrm{C5a}$  & $10^5$    & $10^{-1}$             & $15$  & $100^2\!\times\!\left( 4 \! + \! 2 \! \times \! 30 \right)$   & $11$   & $9,030$  & $730.00$             \\
        $\mathrm{C5b}$  & $10^5$    & $10^{-1/2}$           & $15$  & $100^2\!\times\!\left( 4 \! + \! 2 \! \times \! 30 \right)$   & $11$   & $16,530$ & $227.37$          \\
        $\mathrm{C5c}$  & $10^5$    & $10^{0}$              & $15$  & $100^2\!\times\!\left( 4 \! + \! 2 \! \times \! 30 \right)$   & $11$   & $17,730$ & $70.00$              \\
        $\mathrm{C5d}$  & $10^5$    & $10^{1/4}$            & $15$  & $100^2\!\times\!\left( 4 \! + \! 2 \! \times \! 30 \right)$   & $11$   & $5,530$  & $38.91$           \\
        $\mathrm{C5e}$  & $10^5$    & $10^{1/2}$            & $15$  & $100^2\!\times\!\left( 4 \! + \! 2 \! \times \! 30 \right)$   & $11$   & $10,030$ & $21.54$           \\
        $\mathrm{C5f}$  & $10^5$    & $10^{3/4}$            & $15$  & $100^2\!\times\!\left( 4 \! + \! 2 \! \times \! 30 \right)$   & $11$   & $3,030$  & $12.09$           \\
        $\mathrm{C5g}$  & $10^5$    & $10^{1}$              & $15$  & $100^2\!\times\!\left( 4 \! + \! 2 \! \times \! 30 \right)$   & $11$   & $4,530$  & $6.71$            \\
        $\mathrm{D5}$   & $10^5$    & $\rightarrow \infty$  & $15$  & $100^2\!\times\!4$                                            & $11$   & $3,000$  & $\mathrm{-}$      \\
        \multicolumn{8}{l}{} \\
        $\mathrm{C5cG1}$  & $10^5$    & $10^{0}$            & $1$  & $100^2\!\times\!\left( 4 \! + \! 2 \! \times \! 4 \right)$    & $11$   & $8,530$ & $4.70$              \\
        \textcolor{gray}{$\mathrm{C5c}$} & \textcolor{gray}{$10^5$} & \textcolor{gray}{$10^{0}$} & \textcolor{gray}{$15$} & \textcolor{gray}{$100^2\!\times\!\left( 4 \! + \! 2 \! \times \! 30 \right)$} & \textcolor{gray}{$11$} & \textcolor{gray}{$17,730$} & \textcolor{gray}{$70.00$} \\
        $\mathrm{C5cG30}$  & $10^5$    & $10^{0}$           & $30$  & $100^2\!\times\!\left( 4 \! + \! 2 \! \times \! 45 \right)$   & $11$   & $20,030$ & $140.30$            \\
\end{tabular}
\caption{\justifying{Simulation parameters.
The Prandtl number $\Pr \! = \! 1$ in a horizontally periodic domain of (horizontal) aspect ratio $\Gamma \! = \! 30$ and no-slip conditions at the two solid-fluid interfaces.
The table contains beside the identifier further the Rayleigh number $\Ra$, the thermal diffusivity ratio $\kapparatioIL$, the vertical aspect ratio (or thickness) $\Gs$ of each of the two adjacent solid plates, the total number of spectral elements $N_{e} \! = \!  N_{e,x} \! \times \! N_{e,y} \! \times \! \left( N_{e,z,f} \! + \! 2 \! \times N_{e,z,s} \! \right)$, the polynomial order $N$ on each spectral element, the total simulation runtime $t_r$, and the applied mean temperature drop across each solid plate $\left( \Th - \Tc - 1 \right) \! / 2$.
}}
\label{tab:simulation_parameters}
\end{table}
\let\hp\undefined
\let\hpp\undefined

\rev{Panels (e--h) demonstrate how various azimuthally averaged Fourier energy spectra develop over the course of this gradual aggregation process. 
Here, given $E_{\Phi_{1} \Phi_{2}} := \langle 1/2 \thickspace \Real \left( \hat{\Phi}_{1} \hat{\Phi}_{2}^{*} \right) \rangle_{\phi}$ with the Fourier coefficients $\hat{\Phi} \equiv \hat{\Phi} \left( \bm{k}_{h}, z, t \right)$ and azimuthal angle $\phi$, we include the averaged spectra associated with the temperature field $E_{TT}$, vertical convective heat flux $E_{u_{z} T}$, and vertical velocity $E_{u_{z} u_{z}}$ as a function of the absolute horizontal wave number $k_h$. 
As pattern formation progresses, spectral energy shifts towards smaller $k_h$ -- ie. larger structures -- across all of these spectra, thereby underlining the increasing dominance of large-scale flow structures. 
However, in comparison to the Neumann case described in \citet{Vieweg2021} or \citet{Vieweg2023a}, the aggregation process in the CHT set-up may cease before reaching the domain size depending on $\kapparatioIL$. This dependence of the final size of flow structures (as quantified by $\intLS$, see again eq. \eqref{eq:def_integral_length_scale}) and $\kapparatioIL$ will be analysed in more detail in section \ref{sec:Quantitative_analysis_of_convective_flow_patterns}.}

In this work, due to the inclusion of solid sub-domains and thus the two solid-fluid interfaces, we complement the above measures by the dynamically manifesting temperature drop $\dTN$ across the fluid layer. We find that this measure requires similar time scales of convergence with particularly large times observed for moderate $\kapparatioIL \in \left[ 10^{-1}, 10^{1/2} \right]$. Note that we always require the temperature drop across the fluid layer $\dTN \simeq 1$ (so the non-dimensionalisation for equations \eqref{eq:CE} -- \eqref{eq:EE_s} holds) whereas the temperature drop across each solid plate $\left( \Th - \Tc -1 \right) / 2$ depends strongly on $\kapparatioIL$.

Starting from initial conditions as described in section \ref{subsec:Numerical_domain_boundary_and_initial_conditions}, we run each numerical simulation as long as necessary to reach the late statistically stationary regime (probed by $\intLS$, $\Theta_{rms}$, and $\dTN$) and cover the latter for an extended period of time. Table \ref{tab:simulation_parameters} summarises the simulation parameters for all of our simulation runs.

\begin{figure}
\centering
\includegraphics[width = \textwidth]{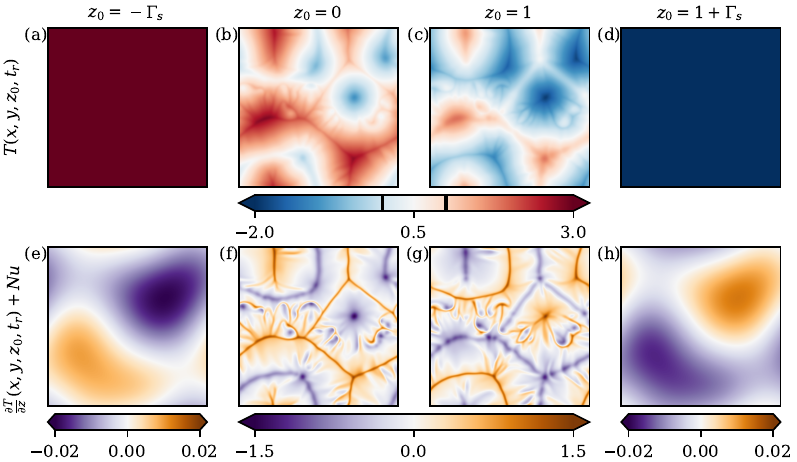}
\caption{\justifying{
Conjugate heat transfer.
In the coupled system, both the temperature (a -- d) and heat flux (e -- h) are coupled at the two solid-fluid interfaces (b, c, f, g) while only the temperature field is controlled at the very bottom (a) and top (d). The respective local heat flux (e, h) is still correlated. 
Here $\Ra = 10^{5}$, $\Gs = 15$, and $\kapparatioIL = 10^{0}$ (i.e., case C4c). 
Note that when $\kapparatioIL \rightarrow \infty$ or $\kapparatioIL \rightarrow 0$, either (b, c) or (f, g) become constant, respectively.
}}
\label{fig:distribution_T_and_dTdz}
\end{figure}

Figure \ref{fig:distribution_T_and_dTdz} underlines the increased complexity of these simulations due to the added solid sub-domains and their coupled interaction with the fluid layer. While we apply Dirichlet-type fixed temperatures at the very top and bottom of the domain -- see panels (a, d) -- the local heat flux at these planes may vary in space and time as visualised in panels (e, h). The coupled heat transfer at the two solid-fluid interfaces implies that we control neither the temperature nor the local heat flux -- see panels (b, c) or (f, g), respectively --, allowing for significantly weaker constraints on the dynamical fluid system.
Interestingly, we find \rev{the vertical temperature gradient fields}
not only to be strongly negatively correlated across the fluid layer (i.e., between planes $z_{0} = \left\{ 0, 1 \right\}$) but also across the entire solid-fluid-solid domain (i.e., between planes $z_0=\left\{- \Gs, 1 + \Gs \right\}$) despite $\Gs = 15$. 
\rev{The temperature fields, on the other hand, appear to be shifted from generally warmer temperatures at $z_0=0$ to colder ones at $z_0=1$ while still showing the footprint of the underlying flow structures.} This underlines the pronounced interplay between solid thermal capacities and the fluid flow.

\subsection{Convective flow patterns for different $\kapparatioIL$}
\label{sec:Convective_flow_patterns_for_different_kapparatio}

We start by resembling the classical, well-known Neumann- \citep{Vieweg2021, Vieweg2023a} and Dirichlet-type \citep{Pandey2018, Vieweg2021} thermal boundary conditions using our CHT set-up subjected to the extreme $\kapparatioIL = \left\{ 10^{-6}, 10^{6} \right\}$. Appendix \ref{sec:Appendix_convergence} contrasts the resulting flow structures -- which are commonly distinguished as supergranules \citep{Vieweg2021} and turbulent superstructures \citep{Pandey2018} (or spiral defect chaos for this lower $\Ra$), respectively -- and confirms a convergence of the flow for plate-less and CHT configurations. 

\begin{figure}
\centering
\includegraphics[width = \textwidth]{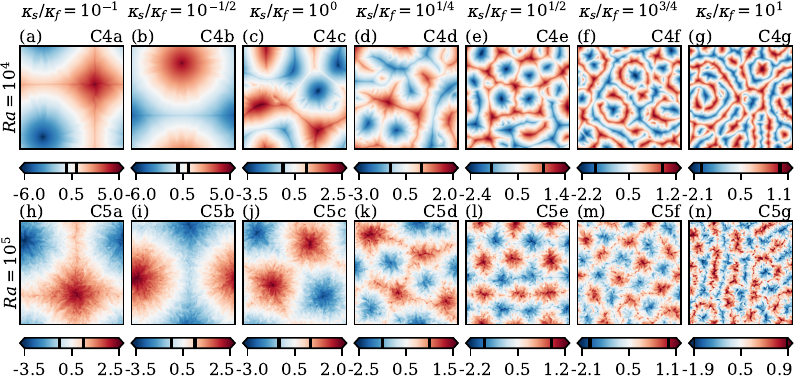}
\caption{\justifying{
Non-linear pattern formation.
Worse solid thermal conductors (relative to the fluid) lead to the formation of larger flow structures.
Here we visualise the instantaneous temperature fields $T \left( x, y, z = 0.5, t = t_{\textrm{r}} \right)$ given $\Gs = 15$. Identifiers for the runs (see the top-right of each panel) are listed in table \ref{tab:simulation_parameters}.
}}
\label{fig:non_linear_pattern_formation}
\end{figure}

Bridging the gap between these previously studied idealised conditions, figure \ref{fig:non_linear_pattern_formation} visualises snapshots of our simulations applying natural thermal boundary conditions covering the range $10^{-1} \leq \kapparatioIL \leq 10^{1}$. Our simulations show that smaller $\kapparatioIL$ lead gradually to increased flow structures given a sufficiently extended domain. In our case, the growth of the flow structures is limited by the numerically finite horizontal domain size $\Gamma = 30$ at $\kapparatioIL = 10^{-1/2}$ and below. Although we study a limited range of $\kapparatioIL$ around unity only, it is sufficient to indicate the clear convergence towards either the supergranules or turbulent superstructures. 
This gradual transition from one of the latter to the other suggests to cover both of them under the umbrella term of long-living large-scale flow structures. 
Our numerical results for $\Ra \gg \Racrit$ are in line with the general, monotonic trend suggested by the linear stability analysis from section \ref{sec:Linear_stability_analysis_of_the_coupled_system}. Independently of $\kapparatioIL$, we find that a larger $\Ra$ introduces stronger turbulence on smaller scales including the granular scale \citep{Vieweg2021, Vieweg2023a}.

\subsection{Quantitative analysis of convective flow patterns}
\label{sec:Quantitative_analysis_of_convective_flow_patterns}

\begin{table}
\centering
\begin{tabular}{@{\hskip 0mm} l @{\hskip 2.0mm} c @{\hskip 2.0mm} c @{\hskip 2.0mm} c @{\hskip 2.0mm} c @{\hskip 2.0mm} c @{\hskip 2.0mm} c @{\hskip 2.0mm} c @{\hskip 2.0mm} c @{\hskip 0mm}}
Identifier          & $t_{ss}$    & $\dTN$      & $\maxDhT$  & $\textrm{std} \left( T \right) $                                            & $\Nu$   & $\Re$  & $\intLS$ & $\rev{\Nug}$     \\ [3pt]
$\mathrm{N4}$   & $2,000$   & $0.316 \pm 0.001$ & $3.04 \pm 0.02$  & $0.55 \pm 0.00$   & $3.17 \pm 0.01$ & $22.82 \pm 0.02$  & $29.86 \pm 0.00$ & $\rev{-}$     \\
$\mathrm{C4N}$  & $2,250$   & $1.001 \pm 0.001$ & $9.66 \pm 0.04$  & $1.77 \pm 0.00$   & $3.16 \pm 0.01$ & $22.79 \pm 0.01$  & $29.86 \pm 0.00$ & $\rev{9.50 \pm 0.01}$     \\
$\mathrm{C4a}$  & $2,250$   & $0.996 \pm 0.001$ & $9.12 \pm 0.02$  & $1.68 \pm 0.00$   & $2.98 \pm 0.00$ & $21.85 \pm 0.00$  & $29.86 \pm 0.00$ & $\rev{8.95 \pm 0.01}$     \\
$\mathrm{C4b}$  & $2,000$   & $1.008 \pm 0.001$ & $8.16 \pm 0.01$  & $1.53 \pm 0.00$   & $2.68 \pm 0.00$ & $20.20 \pm 0.00$  & $29.85 \pm 0.00$ & $\rev{8.04 \pm 0.01}$     \\
$\mathrm{C4c}$  & $2,000$   & $1.000 \pm 0.006$ & $3.74 \pm 0.06$  & $0.72 \pm 0.01$   & $2.43 \pm 0.01$ & $18.61 \pm 0.00$  & $25.68 \pm 1.10$ & $\rev{7.30 \pm 0.03}$     \\
$\mathrm{C4d}$  & $2,000$   & $0.995 \pm 0.005$ & $2.12 \pm 0.07$  & $0.41 \pm 0.01$   & $2.37 \pm 0.01$ & $18.21 \pm 0.01$  & $15.54 \pm 0.78$ & $\rev{7.11 \pm 0.03}$     \\
$\mathrm{C4e}$  & $2,000$   & $0.996 \pm 0.004$ & $1.19 \pm 0.03$  & $0.23 \pm 0.00$   & $2.31 \pm 0.01$ & $17.85 \pm 0.01$  & $~~9.46 \pm 0.29$ & $\rev{6.92 \pm 0.02}$     \\
$\mathrm{C4f}$  & $2,000$   & $1.002 \pm 0.004$ & $0.64 \pm 0.02$  & $0.12 \pm 0.00$   & $2.30 \pm 0.01$ & $17.81 \pm 0.01$  & $~~6.86 \pm 0.17$ & $\rev{6.89 \pm 0.02}$     \\
$\mathrm{C4g}$  & $2,000$   & $1.010 \pm 0.002$ & $0.34 \pm 0.01$  & $0.07 \pm 0.00$   & $2.28 \pm 0.01$ & $17.67 \pm 0.01$  & $~~5.74 \pm 0.11$ & $\rev{6.83 \pm 0.01}$     \\
$\mathrm{C4D}$  & $2,000$   & $1.000 \pm 0.000$ & $0.00 \pm 0.00$  & $0.00 \pm 0.00$   & $2.23 \pm 0.01$ & $17.36 \pm 0.01$  & $~~4.57 \pm 0.10$ & $\rev{2.23 \pm 0.01}$     \\
$\mathrm{D4}$  & $2,000$    & $1$               & $-$               & $-$               & $2.24 \pm 0.02$ & $17.54 \pm 0.01$  & $~~4.42 \pm 0.13$ & $\rev{-}$     \\
\multicolumn{8}{l}{} \\
$\mathrm{N5}$   & $2,000$   & $0.198 \pm 0.001$ & $0.93 \pm 0.02$  & $0.16 \pm 0.00$   & $5.04 \pm 0.03$ & $77.36 \pm 0.53$  & $29.83 \pm 0.01$ & $\rev{-}$     \\
$\mathrm{C5a}$  & $2,000$   & $1.002 \pm 0.004$ & $4.51 \pm 0.06$  & $0.78 \pm 0.00$   & $4.85 \pm 0.05$ & $76.06 \pm 0.43$  & $29.83 \pm 0.01$ & $\rev{14.56 \pm 0.07}$     \\
$\mathrm{C5b}$  & $2,000$   & $1.007 \pm 0.004$ & $4.41 \pm 0.06$  & $0.77 \pm 0.00$   & $4.76 \pm 0.05$ & $75.11 \pm 0.44$  & $29.83 \pm 0.01$ & $\rev{14.28 \pm 0.06}$     \\
$\mathrm{C5c}$  & $2,000$   & $1.004 \pm 0.003$ & $3.09 \pm 0.05$  & $0.55 \pm 0.01$   & $4.65 \pm 0.04$ & $72.90 \pm 0.36$  & $29.43 \pm 0.06$ & $\rev{13.94 \pm 0.05}$     \\
$\mathrm{C5d}$  & $2,000$   & $1.002 \pm 0.002$ & $1.99 \pm 0.06$  & $0.33 \pm 0.00$   & $4.60 \pm 0.04$ & $72.50 \pm 0.32$  & $17.52 \pm 0.92$ & $\rev{13.81 \pm 0.04}$     \\
$\mathrm{C5e}$  & $2,000$   & $1.004 \pm 0.002$ & $1.33 \pm 0.03$  & $0.23 \pm 0.00$   & $4.52 \pm 0.04$ & $71.42 \pm 0.33$  & $10.52 \pm 0.26$ & $\rev{13.57 \pm0.04 }$     \\
$\mathrm{C5f}$  & $2,000$   & $1.004 \pm 0.001$ & $0.85 \pm 0.02$  & $0.15 \pm 0.00$   & $4.49 \pm 0.04$ & $70.65 \pm 0.30$  & $~~9.23 \pm 0.24$ & $\rev{13.51 \pm 0.04}$     \\
$\mathrm{C5g}$  & $2,000$   & $1.005 \pm 0.001$ & $0.50 \pm 0.01$  & $0.09 \pm 0.00$   & $4.45 \pm 0.03$ & $70.01 \pm 0.27$  & $~~6.86 \pm 0.17$ & $\rev{13.35 \pm 0.03}$     \\
$\mathrm{D5}$  & $2,000$    & $1$               & $-$               & $-$               & $4.34 \pm 0.02$ & $68.60 \pm 0.25$  & $~~4.60 \pm 0.09$ & $\rev{-}$     \\
\multicolumn{8}{l}{} \\
$\mathrm{C5cG1}$  & $2,000$   & $1.003 \pm 0.003$ & $1.84 \pm 0.05$  & $0.29 \pm 0.00$   & $4.69 \pm 0.04$ & $72.96 \pm 0.34$  & $15.11 \pm 0.45$ & $\rev{14.06 \pm 0.05}$     \\
\textcolor{gray}{$\mathrm{C5c}$}  & \textcolor{gray}{$2,000$}   & \textcolor{gray}{$1.004 \pm 0.003$} & \textcolor{gray}{$3.09 \pm 0.05$}  & \textcolor{gray}{$0.55 \pm 0.01$}   & \textcolor{gray}{$4.65 \pm 0.04$} & \textcolor{gray}{$72.90 \pm 0.36$}  & \textcolor{gray}{$29.43 \pm 0.06$} & $\rev{13.94 \pm 0.05}$     \\
$\mathrm{C5cG30}$  & $2,000$   & $0.997 \pm 0.003$ & $3.65 \pm 0.07$  & $0.63 \pm 0.00$   & $4.69 \pm 0.04$ & $72.47 \pm 0.34$  & $29.71 \pm 0.03$ & $\rev{14.07 \pm 0.05}$     \\
\end{tabular}
\caption{\justifying{Thermal and global characteristics 
of the direct numerical simulations listed in table \ref{tab:simulation_parameters}. The table contains the analysis time interval in the statistically stationary regime $t_{ss}$ (being part of $t_r$ and situated at its end), the mean temperature difference across the fluid layer $\dT$, the maximum instantaneous temperature difference $\maxDhT$ as an average of the values from the two solid-fluid interfaces, the instantaneous standard deviation of the temperature field at these interfaces $\textrm{std} \left( T \right)$ (again as average over both interfaces), the global Nusselt number $\Nu$, Reynolds number $\Re$, the pattern size as quantified by the integral length scale $\intLS$, \rev{as well as the global CHT Nusselt number $\Nug$}. All characteristics are provided as their time-averaged value together with the corresponding standard deviation. \rev{Note that the error of $\Nug$ has been obtained by calculating the combined uncertainty, regarding $\Nu$ and $\dTN$ as uncorrelated since the correlation coefficient is unknown.}
}}
\label{tab:simulation_outcome_characteristics}
\end{table}

We proceed by quantifying selected aspects of our flow structures as well as their induced statistical properties. 
Firstly, we measure the strength of induced thermal inhomogeneities at the two solid-fluid interfaces based on both the instantaneous maximum horizontal temperature difference 
\begin{equation}
\label{eq:def_max_Delta_h_T}
\maxDhT \left( t \right) :=  \max_{x, y} \left( T \right) - \min_{x, y} \left( T \right)
\end{equation}
and the standard deviation $\stdT$ for $T \in \left\{\Tb, \Tt \right\}$.
This is complemented by the induced global momentum transfer as measured by the Reynolds number \citep{Scheel2017}
\begin{equation}
\label{eq:def_Reynolds_number}
\Re (t)
:= \sqrt{\frac{\Ra}{\Pr}} ~ u_{rms} \qquad \textrm{with} \qquad u_{rms} := \sqrt{ \left\langle \bm{u}^{2} \right\rangle_{V} } .
\end{equation}
Thirdly, we measure the size or horizontal extent of flow structures based on the integral length scale $\intLS$ as defined in equation \eqref{eq:def_integral_length_scale}.

Table \ref{tab:simulation_outcome_characteristics} summarises the temporal averages and associated temporal standard deviations for all our simulations. Note that this analysis covers an extended period $t_{ss}$ of the late statistically stationary regime of pattern formation rather than a single snapshot.

We remark that reaching exactly $\dTN = 1$ is not possible in a CHT set-up -- although being assumed in our non-dimensionalisation, see again section \ref{subsec:Governing_equations} --, affecting in turn the Rayleigh number as defined in equation \eqref{eq:def_Rayleigh_number_Prandtl_number} and thus also $\Nu$ and $\Re$ from equations \eqref{eq:def_Nusselt_number} and \eqref{eq:def_Reynolds_number}, respectively. 
Therefore, the achieved Nusselt and Reynolds numbers have been corrected by this error $\delta T = \dTN - 1$ according to
\begin{equation}
    \label{eq:EA_Re_Nu_corr}
    \Re = \frac{\Re^a}{\sqrt{\dTN}}= \frac{\Re^a}{\sqrt{1+ \delta T}} \sim \delta T^{-1/2} 
    \qquad \textrm{and} \qquad
    \Nu = \frac{\Nu^a}{\dTN}  =  \frac{\Nu^a}{1 + \delta T}  \sim \delta T^{-1},
\end{equation}
where $\Phi^a$ denotes the achieved values by the simulation under presence of $\dTN \neq 1$. 

\rev{In addition to the typical definition of the global Nusselt number from eq. \ref{eq:def_Nusselt_number} -- considering the fluid domain only --, one may also define a kind of CHT Nusselt number $\Nug$ which relates the heat current through the entire CHT set-up (including the solid sub-domains) to the diffusive heat current through the fluid layer}
\begin{subequations}
\label{eq:def_Nu_global}
\begin{align}
    \Nug \left( t \right) :=& 
    \frac{\left\langle \bm{J} \cdot \bm{e}_{z} \right\rangle_{V}}{\langle \bm{J}_{dif,f} \cdot \bm{e}_{z} \rangle_{V_f}} = \frac{\sum_{\Phi} \left \langle \bm{J}_{dif,\Phi} \cdot \bm{e}_{z}\right \rangle_{V_{\Phi}} + \left \langle \bm{u} T_f \cdot \bm{e}_{z}\right \rangle_{V_{f}}}{\langle \bm{J}_{dif,f} \cdot \bm{e}_{z} \rangle_{V_f}} = \\
    =& \frac{{\left \langle u_zT \right \rangle}_{V_f}}{\kappa_f\frac{\Delta T_f}{H}} + 1 + \frac{2}{\Gs} \kapparatio\frac{\Delta T_s}{\Delta T_f} = \Nu \left(t \right)+\frac{2}{\Gs}\kapparatio\frac{\Delta T_s}{\Delta T_f},
\end{align}
\end{subequations}
\rev{where $\Phi=\left\{f,sb,st\right\}$ and $\Delta T_{s,f}$ are the non-dimensional temperature drops across the solid or fluid, respectively (i.e., $\Delta T_f=\dTN$ and $\Delta T_s=\left(T_h-T_c-\Delta T_f\right)\!/2$).
We stress that $\Nug \geq \Nu$, i.e., it represents the latter plus an additional offset that vanishes for $\kapparatioIL \rightarrow \infty$ only.}

\begin{figure}
\centering
\includegraphics[width = \textwidth]{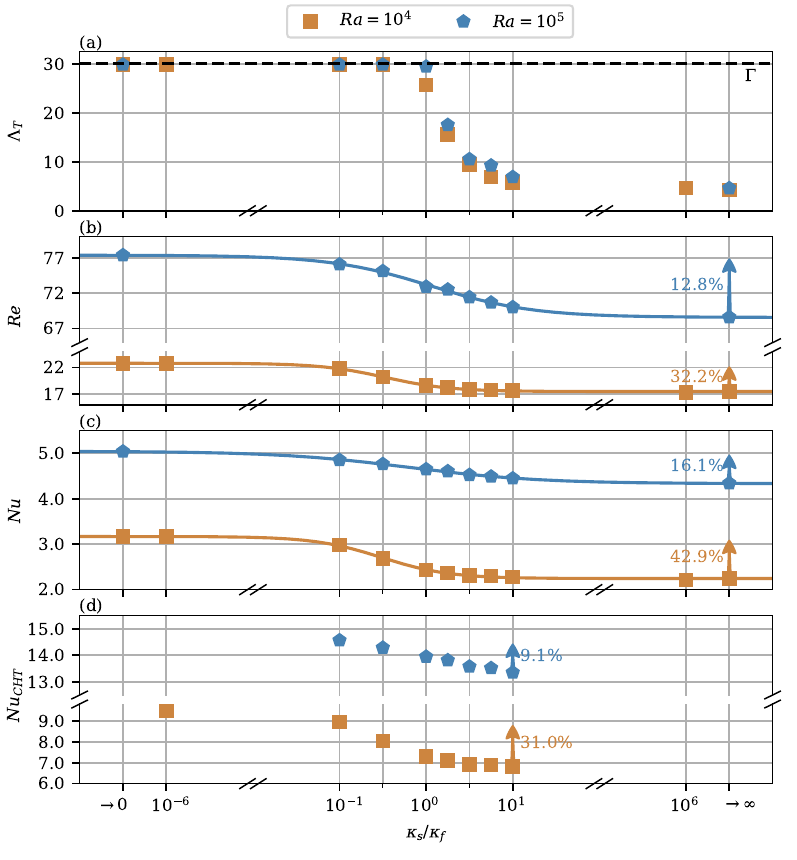}
\caption{\justifying{
Size of flow structures and their induced transport.
The worst solid thermal conductors (relative to the fluid) -- and thus the largest flow structures -- induce strongest turbulence and the biggest global heat transfer. 
Solid lines indicate regressions of the data points based on a hyperbolic tangent function with parameters described by table \ref{tab:regression_parameters_Nu_Re}.
$\Gs = 15$ for all data.
\rev{Note that, in panel (d), $\Nug ( \Ra = 10^{4}, \kapparatioIL = 10^{6} ) = 2.23$ lies beyond the axis limits.}
}}
\label{fig:trends_intLS_Re_Nu}
\end{figure}

Figure \ref{fig:trends_intLS_Re_Nu} illustrates the overall trends of the size of flow structures and their induced momentum and heat transfer for our two main series of simulations across different $\kapparatioIL$. 
First, see panel (a), we find that the qualitative impression from figure \ref{fig:non_linear_pattern_formation} is clearly supported by $\intLS \left( \kapparatioIL \right)$. While such a growth of flow structures is in line with our linear stability analysis from section \ref{sec:Linear_stability_analysis_of_the_coupled_system}, we note that we would expect to see a further but still finite growth of turbulent flow structures for $\kapparatioIL \in \left( 10^{-1}, 10^{-1/2} \right)$ if we provided a sufficiently larger numerical domain. 

As shown by table \ref{tab:simulation_outcome_characteristics}, larger flow structures -- or in other words, smaller $\kapparatioIL$ -- naturally induce stronger thermal heterogeneities at the solid-fluid interfaces. This relaxes the bounds on the temperature field (that is felt by the fluid) and allows thus for a stronger local volumetric forcing in the Navier-Stokes equation \eqref{eq:NSE}. As a result, we find an increased global transfer of momentum as shown by $\Re \left( \kapparatioIL \right)$ in panel (b). Consequently and similarly, also the global heat transfer across the fluid layer is enhanced as measured by $\Nu \left( \kapparatioIL \right)$ in panel (c).
These trends are in line with previous results by \citet{Vieweg2023a} for the extreme Neumann and Dirichlet cases.
\rev{In addition, see panel (d), $\Nug \left( \kapparatioIL \right)$ offers trends similar to $\Nu \left( \kapparatioIL \right)$.}

Starting from the Dirichlet case and moving towards the Neumann case, we find that both $\Re$ and $\Nu$ experience substantial increases of up to $32 \%$ and $43 \%$, respectively. Albeit these relative changes decrease with increasing $\Ra$ due to the increased turbulent mixing, the absolute change of $\Re$ seems still to increase. This underlines that thermal boundary conditions may even affect scaling laws such as $\Nu \sim \Ra^{\gamma}$ \citep{Plumley2019, Vieweg2023a} \rev{in certain ranges}.
\rev{In more detail, the trends in global heat transfer across the fluid layer of our 3-dimensional CHT simulations in a square $\Gamma = 30$ domain align qualitatively with the ones of \citet{Johnston2009}, the latter of which compared the Dirichlet and Neumann cases for $\Gamma=2$ at $\Ra \leq 10^{10}$ in a 2-dimensional domain and found that $\Nu$ is increased in the Neumann case for $\Ra \lesssim 10^6$. In contrast, the impact of the thermal boundary conditions vanished for $\Ra > 10^6$. The same trend was observed by \citet{Vieweg2023a} for 3-dimensional domains of square $\Gamma=60$ at $\Ra \lesssim 10^7$ and by \citet{Verzicco2008} in a cylinder of $\Gamma=1/2$ at $\Ra > 10^9$. 
Albeit this evidence may lead one to suspect that the impact of $\kapparatioIL$ on $\Nu$ and $Re$ vanishes for $\Ra \gg 10^5$ due to increased turbulent mixing, further studies at higher $\Ra$ are required to prove it.}

\begin{table}
  \begin{center}
\def~{\hphantom{0}}
  \begin{tabular}{lcccccc}
       $f$    & $\Ra$  & $a$    & $b$   & $c$    & $d$    & $R^2$ \\[3pt]
       $\Nu$  & $10^4$ & -0.463 & 0.558 & ~0.660 & ~2.708 & 0.9985 \\
       $\Nu$  & $10^5$ & -0.357 & 0.286 & ~0.152 & ~4.687 & 0.9982 \\
       $\Re$  & $10^4$ & -2.673 & 0.578 & ~0.630 & 20.158 & 0.9981 \\
       $\Re$  & $10^5$ & -4.402 & 0.369 & -0.055 & 72.969 & 0.9973 \\
  \end{tabular}
  \caption{\justifying{Regression parameters for $\Nu$ and $\Re$. 
  A $\tanh$-fit of the form $f \left(\kapparatioIL, \Ra \right) = a \tanh{ \left[ b  \log{ \left( \kapparatioIL \right) } + c \right] } + d$ is applied to the values in figure \ref{fig:trends_intLS_Re_Nu}. $R^2$ is the coefficient of determination \citep{Wright1921} and underlines the quality of these fits.}}
  \label{tab:regression_parameters_Nu_Re}
  \end{center}
\end{table}

In analogy to section \ref{sec:Linear_stability_analysis_of_the_coupled_system}, we apply hyperbolic tangent fits to our numerical data. While the resulting fits are included in figure \ref{fig:trends_intLS_Re_Nu}, their underlying parameters are provided in table \ref{tab:regression_parameters_Nu_Re} and allow to estimate expected values of $\Re$ and $\Nu$ given $\Pr=1$ and $\Ra=\left\{10^4,10^5\right\}$ under different $\kapparatioIL$. Reminiscent of section \ref{sec:Linear_stability_analysis_of_the_coupled_system}, we observe again an asymmetric behaviour with the inflection point being skewed towards $\kapparatioIL < 10^{0}$.
 

\section{Boundary layer analysis}
\label{sec:Boundary_layer_analysis}

Albeit the global heat and momentum transport through the fluid layer have been quantified in the previous section \ref{sec:Quantitative_analysis_of_convective_flow_patterns}, a detailed knowledge of both its thermal and viscous boundary layers is essential, too. 
On the one hand, the thermal boundary layers account for the majority of the temperature drop across the fluid layer \citep{Scheel2013, Vieweg2021} and thus induce the essential destabilisation of the latter.
On the other hand, the viscous boundary layers are highly dissipative regions \citep{Scheel2013, Vieweg2021} that slow down fluid motions towards the walls and insinuate the strong turbulence present in the adjacent bulk.

While the \textit{thermal boundary layer thickness} is traditionally defined based on the mean (conductive) heat transfer at the top and bottom boundaries \citep{Chilla2012}
\begin{equation}
    \label{eq:definition_delta_T}
    \delta_T = \frac{1}{2 \thickspace \Nu} ,
\end{equation}
there is no equivalent transfer of momentum at these planes into the fluid. 
Instead, the viscous boundary layer is generated by randomly-oriented patches of shear flow and usually lacks a mean flow \citep{Samuel2024}. This structure suggests an alternative measure of the boundary layer thickness based on the full and horizontal velocity fluctuation profiles
\begin{equation}
    \label{eq:definition_u_rms_uh_rms}
    u_{rms} \left( z \right) = \sqrt{ \left\langle u_x^2 + u_y^2 +u_z^2 \right\rangle_{A,t}} \quad \textrm{and} \quad u_{rms}^h \left( z \right) = \sqrt{ \left\langle u_x^2 + u_y^2 \right\rangle_{A,t}}
\end{equation}
where the maxima mark the \textit{viscous fluctuation thicknesses} $\delta_{u,rms}$ and $\delta_{u,rms}^h$, respectively. 
Equivalently, a similar definition for the \textit{thermal fluctuation thickness} $\delta_{\Theta, rms}$ is given by the maximum of the temperature fluctuation profile
\begin{equation}
    \label{eq:Theta_rms_z}
    \Theta_{rms} \left( z \right) = \sqrt{\left\langle \Theta^2 \right\rangle_{A,t}}
\end{equation}
which has been found to strongly correlate with $\delta_T$ \citep{Long2020}. Note the difference between this profile of $\Theta_{rms}$ and its global quantity defined in equation \eqref{eq:Theta_rms_t}.

We prepare this analysis of thermal and viscous boundary layers by increasing the spatial resolution of our simulations: After $t_{r}$ (see table \ref{tab:simulation_parameters}), we change the vertical number of spectral elements within the fluid layer from $4$ to $6$ -- leading to at least $14$ grid points within a boundary layer -- and relax the flow onto this new grid for $50 \tauf$ before analysing the subsequent $100 \tauf$. 
Moreover, we rescale the temperature field for the following post-processing based on the original $\dTN$ \citep{Vieweg2023a, Vieweg2024a} for improved comparability between the Neumann and remaining cases.

\begin{figure}
\centering
\includegraphics[width = \textwidth]{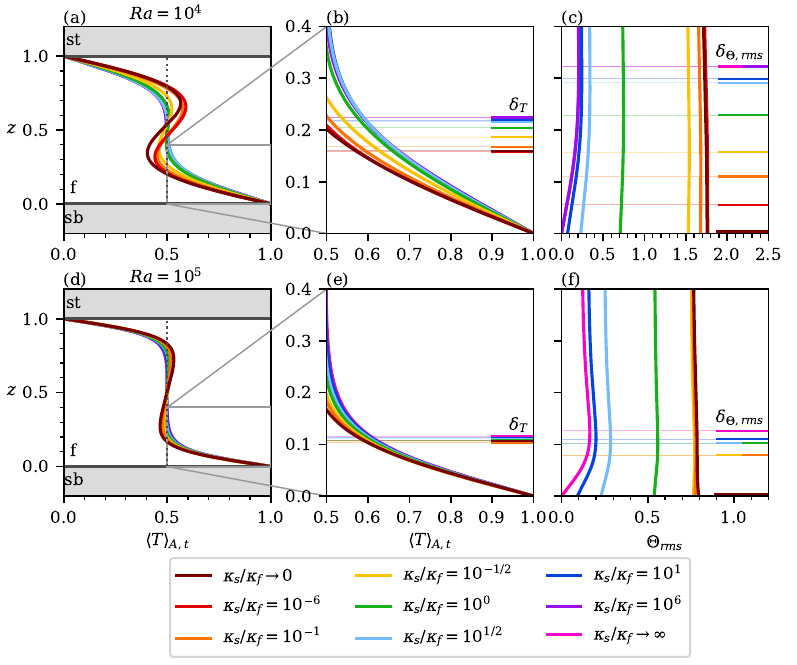}
\caption{\justifying{Thermal boundary layer analysis.
Although (a, d) a glimpse at the entire vertical profiles shows only little variation of them with $\kapparatioIL$, a closer look at the bottom region for both the planar (b, e) average and (c, f) variation reveals a more pronounced dependence of appropriately defined boundary layer thicknesses $\delta_{T}$ and $\delta_{\Theta, rms}$ (as indicated by the horizontal lines).
Note that we exploit the re-scaled temperature field for this analysis and $\Gs = 15$ for all data.
}}
\label{fig:BLA_T}
\end{figure}

Figure \ref{fig:BLA_T} illustrates the mean temperature profiles across our different $\kapparatioIL$ for $\Ra=10^4$ in panels (a, b) and $\Ra=10^5$ in panels (d, e). 
Bridging the gap between idealised thermal boundary conditions reported in \citep{Pandey2022, Vieweg2023a, Vieweg2024a}, our CHT set-up exhibits at this intermediate $\Pr = 1$ an increased tendency for a manifestation of a (weak) stable stratification in the bulk of the fluid layer when decreasing $\kapparatioIL$. This stratification is the result of weakly mixing thermal plumes that detach and shoot deep into or even through the bulk \citep{Vieweg2024a} and might be supported by an increased size and thus large-scale organisation of the flow structures.
As the classical thermal boundary layer thickness $\delta_{T}$ is directly linked to $\Nu$, we find that $\delta_{T}$ decreases with decreasing $\kapparatioIL$. As $\Nu$ tends to become more independent of $\kapparatioIL$ for larger $\Ra$ (see again figure \ref{fig:trends_intLS_Re_Nu}), $\delta_{T}$ converges as well.
As an alternative to this classical measure, panels (c, f) visualise the thermal fluctuation thicknesses $\delta_{\Theta, rms}$. Given natural thermal boundary conditions that are more similar to the Dirichlet case, $\delta_{T}$ and $\delta_{\Theta, rms}$ converge as $\Ra$ is increased \citep{Samuel2024}. However, for conditions more similar to the Neumann case with smaller values of $\kapparatioIL$, the definition of $\delta_{\Theta, rms}$ loses significance as worse solid thermal conductors imply quicker relaxations of thermal perturbations in the fluid and thus a shift of these peaks of variance closer to (or even into) the solids. 

\begin{figure}
\centering
\includegraphics[width = \textwidth]{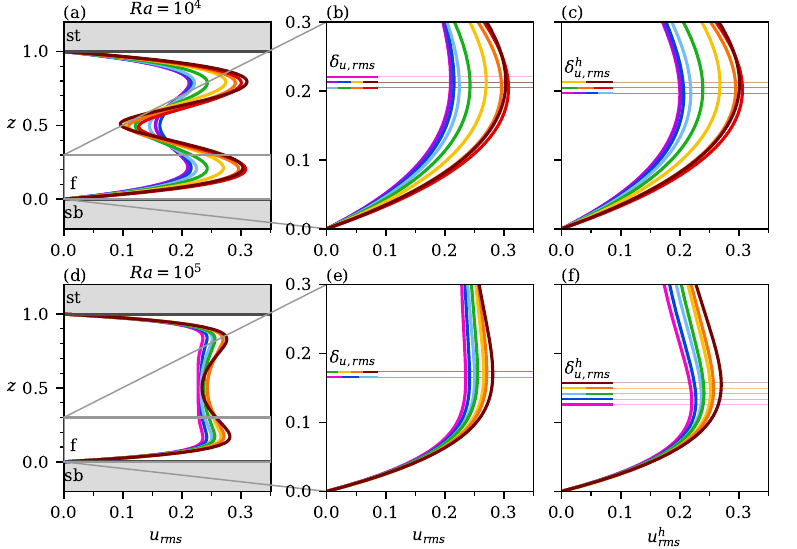}
\caption{\justifying{Viscous boundary layer analysis.
While (a, d) the entire vertical profiles highlight the presence of dominant horizontally extended flow structures in particular for smaller $\kapparatioIL$ and $\Ra$, these profiles' variation allows to derive and contrast appropriately defined boundary layer thicknesses $\delta_{u \textrm{, rms}}$ (as indicated by the horizontal lines) for both the full as well as only the horizontal velocity field.
Note that $\Gs = 15$ for all data and the colour encoding coincides with figure \ref{fig:BLA_T}.
}}
\label{fig:BLA_u}
\end{figure}

Figure \ref{fig:BLA_u} moves the focus of the boundary layer analysis to the velocity field. As shown across the entire fluid domain in panels (a, d), we observe pronounced peaks in the fluctuation profiles especially for smaller $\kapparatioIL$, i.e., for more pronounced large-scale organisations of the flow.
Especially panel (a) indicates the presence of dominant, horizontally extended convection rolls that offer strong horizontal velocities, see also panels (b, c).
Interestingly, despite the very different amplitudes of these profiles across $\kapparatioIL$, the viscous boundary layer thickness is very similar and varies only weakly with $\Ra$ (and thus $\Re$). In other words, the viscous boundary layer thickness varies less with both $\kapparatioIL$ and $\Ra$ than the thermal one. As a result, the ratio $\delta_{u,rms}/\delta_{\Theta,rms}$ tends to increase for larger $\Ra$ \citep{Samuel2024}.
However, our data suggests that it is not solely $\Re$ that governs the thickness of the viscous boundary layer: for fixed $\Ra$, our $\Re$ is larger for smaller $\kapparatioIL$ even though both $\delta_{u,rms}$ and $\delta_{u,rms}^h$ exhibit the opposite trend. This suggests that long-living large-scale flow structures play a crucial role even for viscous boundary layers.


\section{The impact of the plate thickness}
\label{sec:The_impact_of_the_plate_thickness}

\subsection{Choice of the vertical aspect ratio $\Gs$}
\label{sec:Choice_of_the_vertical_aspect_ratio}

\begin{figure}
\centering
\includegraphics[width = \textwidth]{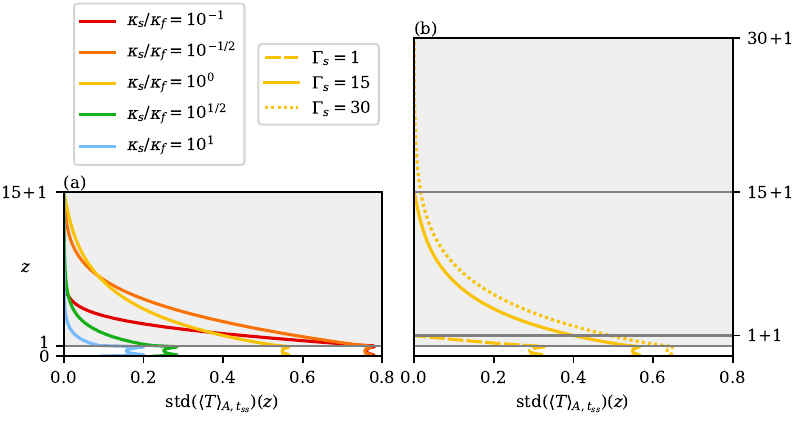}
\caption{\justifying{
Relaxation of turbulent flow-induced thermal perturbations across the solid plates. 
(a) Given $\Gs = 15$ (at $\Ra = 10^{5}$), the relaxation is slowest close to a unity ratio $\kapparatioIL = 10^{0}$. 
Even for this critical case, (b) the situation has mostly converged to that with plates of even twice the thickness. In contrast, thinner plates with $\Gs = 1$ impact the temperature field at the solid-fluid interfaces strongly.
The situation is symmetric for $- \Gs \leq z \leq 0$.
}}
\label{fig:thermal_perturbations_with_GammaS}
\end{figure}

In all our simulations discussed so far, a vertical aspect ratio of $\Gs=15$ has been used. This deliberate choice is based on a diffusion time argument: Temperature differences are supposed to relax more quickly in the horizontal than in vertical direction, 
\begin{equation}
    \label{eq:diffusion_times_hv}
    \tau_{\kappa,s,v}\stackrel{!}{\geq}\tau_{\kappa,s,h} \quad \textrm{with} \quad \tau_{\kappa}=\frac{L_{ch}^2}{\kappa} ,
\end{equation}
promoting a weak to negligible footprint of the applied thermal boundary conditions on the solid-fluid interfaces. 
Given our laterally periodic domain, the largest structure satisfying this condition offers a horizontal extend of $L_{ch,h}=L/2=\Gamma H/2$. Hence, condition \eqref{eq:diffusion_times_hv} turns into
\begin{equation}
    \label{eq:diffusion_times_comp}
    \frac{H_s^2}{\kappa_s} \stackrel{!}{\geq} \frac{L^2/4}{\kappa_s} = \frac{\Gamma^2H^2}{4\kappa_s} \Leftrightarrow \Gs \geq \frac{\Gamma}{2} 
\end{equation}
and a domain of $\Gamma=30$ should offer (at least) $\Gs=15$.

Figure \ref{fig:thermal_perturbations_with_GammaS} scrutinises our condition \eqref{eq:diffusion_times_comp} by plotting and contrasting vertical profiles of the standard deviation in the temperature field $\stdT$ across all $\kapparatioIL$ in panel (a). We find that the propagation of thermal inhomogeneities into the solid plates is asymmetric with respect to $\kapparatioIL$. Interestingly, thermal perturbations relax most slowly in case of $\kapparatioIL = 10^{0}$ despite its weaker thermal inhomogeneities at the solid-fluid interface compared to $\kapparatioIL = \left\{ 10^{-1}, 10^{-1/2} \right\}$ (see also again table \ref{tab:simulation_outcome_characteristics}). 

In order to probe the impact of our applied thermal boundary condition at $z = 1 + \Gs$ (see equation \eqref{eq:Tc}), we proceed by varying $\Gs$ given a fixed $\kapparatioIL = 10^{0}$. Panel (b) contrasts two additional simulations of $\Gs = \left\{ 1, 30 \right\}$ with the previous case.
On the one hand, we find that thin plates of $\Gs=1$ result in significant decreases of the thermal inhomogeneities and size of the resulting flow structures compared to $\Gs = 15$, see also again table \ref{tab:simulation_outcome_characteristics}. This suggests that temperature differences in the horizontal direction are not sufficiently relaxed and that the underlying thermal boundary conditions leave a considerable footprint on the solid-fluid interfaces. 
On the other hand, even thicker plates of $\Gs=30$ do not seem to provide a significant benefit. Despite somewhat larger inhomogeneities at the solid-fluid interfaces, the vertical profile underlines that the relaxation of thermal inhomogeneity is barely altered despite a doubling of the plate thickness. 
Hence, we conclude that the choice of $\Gs = 15$ based on condition \eqref{eq:diffusion_times_comp} for our main production simulations has been appropriate.

\subsection{Linear stability analysis for different plate thicknesses $\Gs$}
\label{sec:Linear_stability_analysis_for_different_plate_thicknesses}

\begin{figure}
\centering
\includegraphics[width = \textwidth]{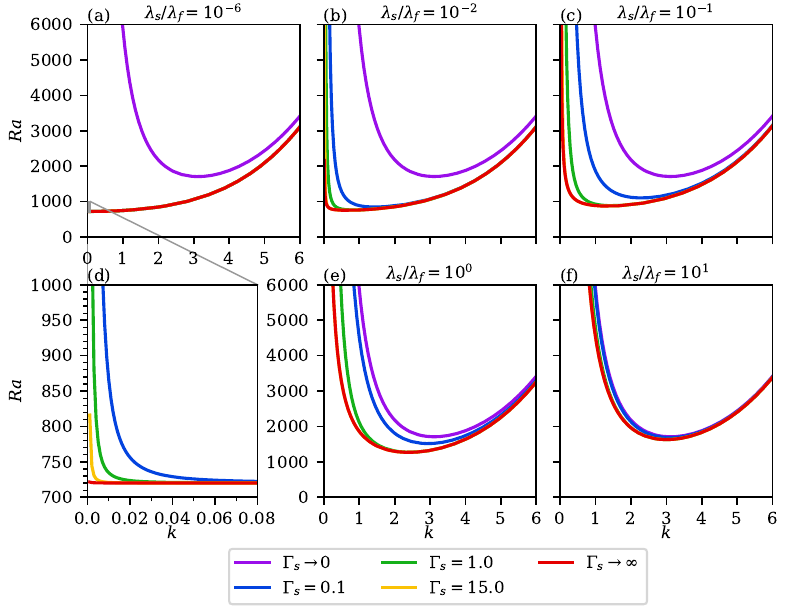}
\caption{\justifying{
Neutral stability across varying $\Gs$.
While vertically infinitely extended plates are already resembled at $\Gs \gg 1$, thinner plates $\Gs \lesssim 1$ impact the system significantly and stabilise the layer successively. 
Worse solid thermal conductors (relative to the fluid) are more strongly affected by this stabilisation; contrast therefore in particular panels (c, f) which are symmetrically spaced around $\lambdaratioIL = 10^{0}$.
}}
\label{fig:neutral_stability_with_GammaS}
\end{figure}

Similar to the turbulent flow at $\Ra \gg \Racrit$, we expect the onset of convection to be affected by the thickness of the solid plates $\Gs$. 
As shown in appendix \ref{sec:Appendix_LSA}, we extend the work of \citet{Hurle1967} (who considered $\Gs \rightarrow \infty$) by respecting the plate thickness via the $-\lambdaratioInvIL \tanh \left(k \Gs \right)$ term in the solution of the linear stability of the system. 

Figure \ref{fig:neutral_stability_with_GammaS} compares various neutral stability curves given constant $\lambdaratioIL$ in different panels. 
On the one hand, we find that the curves (across the different panels) converge for infinitely \textit{small} vertical aspect ratios $\Gs \rightarrow 0$ -- independently of $\lambdaratioIL$ -- towards the classical Dirichlet case (i.e., the violet curves are all the same). This case, making any plates obsolete, is certainly influenced by our applied Dirichlet-type boundary conditions at the very top and bottom.
On the other hand, we observe that the curves (in each panel) converge for \textit{large} vertical aspect ratios $\Gs \gg 1$ independently of $\lambdaratioIL$. In other words, there is practically no difference between $\Gs=15$ and $\Gs \rightarrow \infty$. This substantiates our choice of $\Gs=15$ in section \ref{sec:Choice_of_the_vertical_aspect_ratio}. 
Additionally, we find that neutral stability curves converge for $\lambdaratioIL \rightarrow \left\{ 0, \infty\right\}$ as long as $\Gs > 0$ and sufficiently large (e.g., $\Gs = 0.1$). Only a zoom towards $k \approx 0$, see panel (d), indicates the gradual convergence of the critical wave number for the Neumann case as one of these two idealised thermal boundary conditions. For such small, yet positive $\lambdaratioIL$, the solid plates conduct heat way worse than the fluid, and so even very thin plates of $\Gs=0.1$ effectively act insulating. This may have important implications for any coating of a wall in laboratory convection set-ups \citep{Schindler2022, Schindler2023, Wondrak2023}. 

\begin{figure}
\centering
\includegraphics[width = \textwidth]{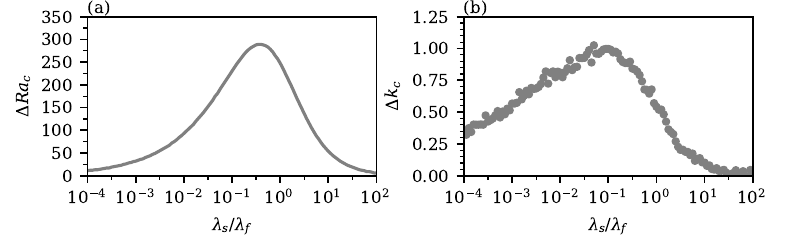}
\caption{\justifying{
Sensitivity of neutral stability on plate thickness for varying $\lambdaratioIL$.
The differences in $\Racrit$ (panel (a)) and corresponding $\kcrit$ (panel (b)) are shown when moving from infinitely thick ($\Gs\rightarrow\infty$) towards very thin plates ($\Gs=0.1$). Decreasing $\Gs$ thus stabilises the layer successively, with the strongest impact near $\lambdaratioIL\approx10^{-1/2}$, not just shifting the onset of convection, but also reducing the initial pattern size at this point.
}}
\label{fig:Delta_Ra_crit_comparison}
\end{figure}

The situation becomes significantly more complex once no limit of either $\Gs$ or $\lambdaratioIL$ is considered. Contrasting panels (b, c, e, f), a certain asymmetry around $\lambdaratioIL=10^0$ can (once again) be noticed. While the neutral stability curves almost coincide for $\lambdaratioIL=10^1$ in panel (f), there is a significant spread for $\lambdaratioIL=10^{-1}$ in panel (c). This begs the following question: Which ratio of thermal conductivities is most sensitive to a variation of $\Gs$? 

In an attempt to quantify the divergence of the neutral stability curves, we measure the differences in $\Racrit$ and $\kcrit$ between the cases of $\Gs = 0.1$ and $\Gs\rightarrow\infty$ via
\begin{align}
    \label{eq:definition_Delta_Ra_crit}
    \Delta \Racrit \left( \lambdaratio \right) &= \Racrit \left( \lambdaratio, \Gs=0.1 \right) - \Racrit \left( \lambdaratio, \Gs \rightarrow \infty\right),  \\
    \label{eq:definition_Delta_k_crit}
    \Delta \kcrit \left( \lambdaratio \right) &= \kcrit \left( \Racrit \left( \lambdaratio, \Gs=0.1 \right) \right) - \kcrit \left( \Racrit \left( \lambdaratio, \Gs \rightarrow \infty\right) \right).
\end{align}
Figure \ref{fig:Delta_Ra_crit_comparison} illustrates these results. 
Note that while $\Delta \Racrit > 0$ implies a stabilisation of the convection layer when decreasing the thickness of the solid plates, $\Delta \kcrit > 0$ indicates a decrease in the size of critical flow structures.
Moreover, we find that both $\Delta \Racrit$ and $\Delta \kcrit$ offer pronounced peaks around $\lambdaratioIL \approx 10^{-0.5}$ and $\lambdaratioIL \approx 10^{-1}$, respectively. Supported by the general asymmetry of these curves, this analysis underlines the complex ramifications an interplay between the solid and fluid domain can exhibit.

\section{Discussion and perspective}
\label{sec:Discussion_and_perspective}

Long-living large-scale flow structures are crucial for an understanding and prediction of convection flows such as in Earth's atmosphere. Previous studies of horizontally extended Rayleigh-Bénard convection focussed on idealised thermal boundary conditions such as constant temperatures (the so-called Dirichlet case) or a constant heat flux (Neumann case) \citep{Pandey2018, Vieweg2021, Vieweg2023a}. However, these conditions reduce the problem to the fluid layer only and thus represent rather idealised cases. In contrast, any natural convection flow is confined by some adjacent matter. 
Using a coupled or conjugate heat transfer (CHT) set-up, this study includes two identical fluid-confining solid plates at the bottom and the top of the fluid layer. The ratio of thermal diffusivities $\kapparatioIL$ between the solids and the fluid represents the key control parameter characterising the (to the perspective of the fluid) resulting thermal boundary conditions. The inclusion of solid sub-domains allows to resemble the Neumann case via $\kapparatioIL \rightarrow 0$ and the Dirichlet case via $\kapparatioIL \rightarrow \infty$, see appendix \ref{sec:Appendix_convergence}. Varying $\kapparatioIL$ across a broad range, this study bridges the gap in between by introducing natural thermal boundary conditions.

Given a Prandtl number $\Pr=1$, (horizontal) aspect ratio $\Gamma=30$ and thickness of the solid plates $\Gs=15$, we have conducted direct numerical simulations subject to varying $\kapparatioIL$ for $\Ra= \left\{10^4, 10^5 \right\}$ under no-slip boundary conditions at the solid-fluid interfaces and periodic boundary conditions concerning the lateral extent of the domain of square horizontal cross-section.
As shown in figure \ref{fig:distribution_T_and_dTdz}, such a coupled system allows for highly complex dynamics of both $T$ and $\partial T / \partial z$ at the various (partly solid-fluid) interfaces based on the manifesting fluid flow.

Varying $\kapparatioIL$ from $\kapparatioIL \gg 10^{0}$ towards $\kapparatioIL \ll 10^{0}$, we found that both structural as well as statistical properties of the flow undergo significant changes: the long-living large-scale flow structures grow, and both their induced momentum and heat transfer are (asymmetrically with respect to $\kapparatioIL$) increased by up to $43 \%$. 
The increase of these measures can be explained by stronger thermal inhomogeneities -- the footprints of these flow structures -- at the solid-fluid interfaces and align with recent results for idealised thermal boundary conditions \citep{Vieweg2021, Vieweg2023a} or strongly asymmetric CHT set-ups \citep{Vieweg2025}.
We observe a gradual transition from turbulent superstructures towards supergranules, underlining the importance of the umbrella term of long-living large-scale flow structures \citep{Vieweg2023a}.

A linear stability analysis of our CHT set-up confirms both the growth of flow structures as well as an increased heat transfer for smaller values of $\lambdaratioIL$. In other words, we observe a monotonic shift of both the critical Rayleigh number $\Racrit$ and critical wavenumber $\kcrit$ when varying $\lambdaratioIL$ between its limits. This implies a change in the supercriticality $\Ra / \Racrit$ given a fixed $\Ra > \Racrit$ and affects thus the induced heat transfer.

We have extended the previous work by \citet{Hurle1967} with respect to two important aspects: 
First, we provided simple relations or regressions for $\kcrit \left( \lambdaratioIL \right)$, $\Racrit \left( \lambdaratioIL \right)$, and $\Racrit \left( \kcrit \right)$ -- see again table \ref{tab:regression_parameters_k_Ra}. This hopefully improves the accessibility of our results.
Second, we investigated the impact of a finite thickness of the solid plates in section \ref{sec:Linear_stability_analysis_for_different_plate_thicknesses}. This allowed us to understand the convergence of the system for the limits $\left\{ \Gs, \lambdaratioIL \right\} \rightarrow \left\{ 0, \infty \right\}$ and find its point of largest susceptibility to a change in the plate thickness (as quantified by $\Delta \Racrit$) at $\lambdaratioIL \approx 10^{-1/2}$. 

This effect of a varying plate thickness $\Gs$ was additionally studied numerically. We found that $\Gs \lesssim 1$ tends to stamp the external thermal boundary condition onto the internal solid-fluid interface, whereas our chosen $\Gs = 15 \gg 1$ represents a fair approximation of $\Gs \rightarrow \infty$. We note at this point that the computational cost of one CHT simulation involving $\Gs = 15$ increases the required wallclock solution time by a factor of roughly $8$ compared to a simulation without solid plates \rev{(for the same amount of non-dimensional time units $\tauf$)}. 
Both this numerical investigation of varying $\Gs$ as well as the corresponding investigation of the linear stability have important implications for the design of laboratory experiments \citep{Foroozani2021, Moller2021, Moller2022, Wondrak2023, Vieweg2025}.

An additional analysis of both the thermal and viscous boundary layers confirmed an increasing ratio between these thicknesses, $\delta_{u,rms} / \delta_{\Theta, rms}$, for increasing $\Ra$ \citep{Samuel2024}, even though we considered two different Rayleigh numbers only. However, our data suggests that long-living large-scale flow structures play an important role in the formation of $\delta_{u,rms}$, so that the latter is not solely governed by $\Re$.

In nature, thermal convection flows offer typical ratios of thermal diffusivities in the range of $\kapparatioIL \approx \left[ 10^{-3}, 10^{-1} \right]$ (see again table \ref{tab:natural_kapparatio_lambdaratio}). Such values are not only clearly between the typically studied cases of $\kapparatioIL \rightarrow 0$ and $\kapparatioIL \rightarrow \infty$, but also shifted towards the only more recently investigated Neumann case. Albeit we have covered corresponding values of $\kapparatioIL$, natural flows exhibit far greater $\Ra$ and potentially smaller $\Pr$ as well as additional mechanisms like rotation \citep{Schumacher2020, Vieweg2022}. Our present study can be seen as an important, yet early step towards understanding these more demanding and sophisticated geophysical and astrophysical systems. 
\rev{Our recent work \citep{Vieweg2025} already started investigating asymmetric top and bottom plates and thermal boundary conditions as present in scientific engineering applications. Together with additional rotation around the vertical axis \citep{Vieweg2022}, this may represent an interesting potential path for future extensions relevant to core-core-mantle or core-ocean-ice configurations found on Earth or icy moons, respectively. Given this last discussion, the present work can define a starting point for further investigations on non-ideal boundary condition effects only.}


\backsection[Funding]
{
P.P.V. is funded by the Deutsche Forschungsgemeinschaft (DFG, German Research Foundation) within Walter Benjamin Programme 532721742. 
M.E. is funded by the Carl Zeiss Foundation with project number P2022-08-006.
The authors gratefully acknowledge the Gauss Centre for Supercomuting e.V. (\href{www.gauss-centre.eu}{\texttt{www.gauss-centre.eu}}) for funding this work by providing computing resources 
through the John von Neumann Institute for Computing (NIC) on the GCS supercomputer JUWELS at Jülich Supercomputing Center (JSC) within project nonbou.
They further acknowledge the computing centre of the Technische Universität Ilmenau for providing access to, as well as computing and storage resources on its compute cluster MaPaCC24.
}

\backsection[Declaration of interests]{
The authors report no conflict of interest.
}

\backsection[Author ORCIDs]{
M. Ettel, https://orcid.org/0009-0008-6962-3587     \\
P. P. Vieweg, https://orcid.org/0000-0001-7628-9902 \\
J. Schumacher, https://orcid.org/0000-0002-1359-4536 \\
}

\backsection[Author contributions]{
P.P.V. and J.S. designed the study. M.E. performed the numerical simulations and processed the generated data. All authors contributed equally in discussing the data and writing the paper.
}

\appendix

\section{Linear stability analysis for the coupled system}
\label{sec:Appendix_LSA}

\subsection{Key idea of the linear stability analysis}
\label{sec:Appendix_LSA_key_idea}

The aim of the linear stability analysis is to find the point of the onset of convection with the fluid being initially at rest. The system is considered stable if perturbations, induced as infinitesimally small fluctuations in the form of planar waves, decay. Vice versa, it is unstable if such perturbations grow over time. The analysis is termed linear since one linearises the governing equations with respect to the (infinitesimally small) perturbations.

As it will turn out, the point of the onset of convection is determined by the critical Rayleigh number $\Racrit$ and is caused by the normal mode of the critical wavenumber $\kcrit$. In order to transition from a linearly stable to a linearly unstable state, the system must pass the so-called marginal state - which is exactly the one defining these critical numbers \citep{Chandrasekhar1981}.

In the following, the linear stability analysis for the CHT case will be performed. 
This will yield not only the neutral stability curves for different thermal conductivity ratios $\lambdaratioIL$, the former of which represent the marginal states via $\Ra \left( k \right)$, but also the critical Rayleigh numbers $\Racrit$ and wave numbers $\kcrit$ as given by their global minimum.

\subsection{Governing equations and boundary conditions}
\label{sec:Appendix_LSA_governing_eq}

We consider the same set-up as shown in figure \ref{fig:schematic_configuration} (b).
As long as $\Ra\!<\!\Racrit$, we are in a non-convective regime and the fluid is at rest. As mentioned in section \ref{subsec:Numerical_domain_boundary_and_initial_conditions} and laid out in more detail in \citep{Vieweg2025}, in both the solids and the fluid the temperature profile will be linear. For further analysis it will be benefitial to use the temperature deviation $\Theta(\bm{x}, t)$. It is the deviation of the actual temperature profile from the linear one (which is present in a non-convective case) for both the solids and the fluid
\begin{align}
    \label{eq:Theta_definition}
    \varTheta_{\Phi}(\bm{x},t) = T_{\Phi}(\bm{x},t) - T_{lin,{\Phi}}(z), \qquad \textrm{with} \qquad \Phi=\{f,sb,st\}.
\end{align}
For this analysis, we use the governing equations \eqref{eq:CE} -- \eqref{eq:EE_s} based on a different non-dimensionalisation. Instead of the free-fall-inertia-balance (favourable for large $\Re$), we consider a scaling based on the viscous diffusion time scale $\tau_{\nu}=H^2 / \nu$ leading to
\begin{align}
    \label{eq:LSA_CE_nondim}
    \Tilde{\nabla} \cdot \Tilde{\bm{u}} &= 0, \\
    \label{eq:LSA_NSE_nondim}
    \frac{\partial \Tilde{\bm{u}}}{\partial t} + ( \Tilde{\bm{u}} \cdot \Tilde{\nabla} ) \Tilde{\bm{u}} &= - \Tilde{\nabla} \Tilde{p} +  \Tilde{\nabla}^2 \Tilde{\bm{u}} + \Ra \Tilde{\Theta} \bm{e}_{z}, \\
    \label{eq:LSA_EE_f_nondim}
     \Pr \frac{\partial \Tilde{\Theta}_{f}}{\partial \Tilde{t}} + \Pr (\Tilde{\bm{u}} \cdot \Tilde{\nabla} ) \Tilde{\Theta}_{f} &= \Tilde{\nabla}^2 \Tilde{\Theta}_{f} + \Tilde{u}_z, \\
    \label{eq:LSA_EE_s_nondim}
    \Pr \kapparatioInv \frac{\partial \Tilde{\Theta}_{s}}{\partial \Tilde{t}} &= \Tilde{\nabla}^2 \Tilde{\Theta}_{s}.
\end{align}
Equations \eqref{eq:LSA_CE_nondim}--\eqref{eq:LSA_EE_s_nondim} are completed by corresponding boundary conditions. In our set-up, mechanical no-slip boundary conditions are applied at both interfaces. Due to the present symmetry in our domain, we place the origin of the $z$-coordinate at the mid-plane of the fluid layer. Together with the continuity equation \eqref{eq:LSA_CE_nondim}, this results in
\begin{align}
    \label{eq:LSA_BC_12}
    \Tilde{\bm{u}} \left(\Tilde{z}=\pm 1/2 \right) &= 0 , \\
    \label{eq:LSA_BC_34}
    \left. \frac{\partial \Tilde{u}_z}{\partial \Tilde{z}}\right|_{\Tilde{z}=\pm 1/2} &= 0.
\end{align}
As the sub-domains are coupled at the solid-fluid interfaces, both the temperatures and heat fluxes must match there which leads to
\begin{align}
    \label{eq:LSA_BC_5}
    \Tilde{\Theta}_{f} \left( \Tilde{z}=1/2 \right) &= \Tilde{\Theta}_{st} \left( \Tilde{z}=1/2 \right), \\
    \label{eq:LSA_BC_6}
    \Tilde{\Theta}_{f} \left( \Tilde{z}=-1/2 \right) &= \Tilde{\Theta}_{sb} \left( \Tilde{z}=-1/2 \right), \\
    \label{eq:LSA_BC_7}
    \left.\frac{\partial \Tilde{\Theta}_{f}}{\partial \Tilde{z}} \right |_{\Tilde{z}=1/2} &= \frac{\lambda_{st}}{\lambda_{f}} \left.\frac{\partial \Tilde{\Theta}_{st}}{\partial \Tilde{z}} \right |_{\Tilde{z}=1/2}, \\
    \label{eq:LSA_BC_8}
    \left. \frac{\partial \Tilde{\Theta}_{f}}{\partial \Tilde{z}} \right |_{\Tilde{z}=-1/2} &= \frac{\lambda_{sb}}{\lambda_{f}} \left. \frac{\partial \Tilde{\Theta}_{sb}}{\partial \Tilde{z}} \right |_{\Tilde{z}=-1/2}. 
\end{align}
We shall omit the tildes in the following for better readability.


\subsection{Perturbation equations}
\label{sec:Appendix_LSA_perturbation_eq}

Next, we apply infinitesimally small perturbations to the system by using a linear combination of basic perturbations. This forms a full set and allows extracting the one at which instability first occurs. All variables will be subject to the perturbation $\phi'$ with $\varepsilon\ll1$ as a perturbation parameter, such that the base state $\Bar{\phi}$ is disturbed via
\begin{align}
    \label{eq:LSA_perturbation_definition}
    \phi = \Bar{\phi} + \varepsilon \phi' \quad \text{for} \quad \phi = \{\bm{u},p,\Theta \}.
\end{align}

Applying the curl to equation \eqref{eq:LSA_NSE_nondim} twice allows to drop the pressure term. At this point, the infinitesimally-small perturbations are applied to all variables. Keeping in mind that the fluid is at rest and the temperature deviation is absent for the base state, implying
\begin{align}
    \label{eq:LSA_Mean_profiles_zero}
    \Bar{\bm{u}}=0, \quad \Bar{\Theta}=0 ,
\end{align}
as well as dividing by $\varepsilon \neq 0$ and dropping all terms of $\textit{O} \left( \varepsilon^2 \right)$ since $\varepsilon \ll 1$ (i.e.,  $\varepsilon^2 \lll 1$) yields the linearised equations
\begin{align}
    \label{eq:LSA_CE_perturbed}
    \nabla \cdot \bm{u}' &= 0, \\
    \label{eq:LSA_NSE_perturbed}
    \frac{\partial (\nabla^2\bm{u}')}{\partial t} &= \nabla^4 \bm{u}' + \Ra \left( \nabla^2 \Theta_{f}' - \nabla \frac{\partial \Theta_{f}'}{\partial z} \right), \\
    \label{eq:LSA_EE_f_perturbed}
    \frac{\partial \Theta_{f}'}{\partial t} &= \nabla^2 \Theta_{f}' + u_z', \\
    \label{eq:LSA_EE_s_perturbed}
    \Pr \kapparatioInv \frac{\partial \Theta_{s}'}{\partial t} &= \nabla^2 \Theta_{s}'.
\end{align}

In a next step, a normal mode ansatz is applied: We assume an infinitely-extended domain in the horizontal directions (in line with our laterally periodic boundary conditions) and use the linear equations \eqref{eq:LSA_CE_perturbed} -- \eqref{eq:LSA_EE_s_perturbed} to superpose plane waves or normal modes which form a complete set of basis functions. These key features allow for linear superposition in the first place and later for an extraction of the wave at which instability occurs first. Such plane waves are given by
\begin{align}
    \label{eq:plane_wave_defintion}
    \phi' = \Hat{\phi}' \left( z \right) e^{ \mathrm{i} \left( k_x x + k_y y \right) + \sigma t },
\end{align}
where $k_{x,y}$ are the wavenumbers in $x$ and $y$ direction, defining the horizontal wavenumber 
\begin{align}
    \label{eq:LSA_k_h}
    k_{h}=k=\sqrt{k_x^2+k_y^2}.
\end{align}
$\sigma \in \mathbb{C}$ is the growth rate. 
Each perturbation $\phi'$ in \eqref{eq:LSA_CE_perturbed} -- \eqref{eq:LSA_EE_s_perturbed} is expressed via \eqref{eq:plane_wave_defintion} in terms of normal modes. 
Exemplary for the continuity equation, this results in
\begin{align}
    \label{eq:LSA_CE_phi}
    \nabla \cdot \bm{u}' \equiv &\left[ \mathrm{i} k_x \Hat{u}_x'\left( z \right) +\mathrm{i}k_y \Hat{u}_y'\left( z \right) + \frac{\partial \Hat{u}_z' \left( z \right) }{\partial z} \right] e^{\mathrm{i}\left( k_x x + k_y y \right) + \sigma t} = 0. 
\end{align}
For abbreviation, we define
\begin{align}
    \label{eq:LSA_abbreviation}
    U:=\Hat{u}_x' \left( z \right), V:=\Hat{u}_y' \left( z \right), W:=\Hat{u}_z' \left( z \right), \Hat{\Theta}' \left( z \right) = \Theta, D\phi:=\frac{\partial \left( \phi \right)}{\partial z},
\end{align}
such that equation \eqref{eq:LSA_CE_phi} translates to simply
\begin{align}
    \label{eq:LSA_CE_final}
    \mathrm{i}k_x U +\mathrm{i}k_y V + D W = 0.
\end{align}
Applying this in a similar fashion to equations \eqref{eq:LSA_NSE_perturbed} in $z$ and \eqref{eq:LSA_EE_f_perturbed} -- \eqref{eq:LSA_EE_s_perturbed}, we obtain
\begin{align}
    \label{eq:LSA_NSE_final}
    \sigma \left( D^2-k^2 \right) W &= \left( D^2 -k^2 \right)^2 W - \Ra k^2 \Theta_{f}, \\
    \label{eq:LSA_EE_f_final}
    \Pr \sigma \Theta_{f} &= \left( D^2 -k^2 \right) \Theta_{f} + W, \\
    \label{eq:LSA_EE_s_final}
    \Pr \kapparatioInv \sigma \Theta_{s} &=  \left( D^2 - k^2 \right) \Theta_{s}.
\end{align} 
On the other hand, the boundary conditions \eqref{eq:LSA_BC_12} -- \eqref{eq:LSA_BC_8} translate to
\begin{align}
    \label{eq:LSA_BC_12_abbrev}
    W\left( z=\pm 1/2 \right) &= 0, \\
    \label{eq:LSA_BC_34_abbrev}
    DW |_{ z=\pm 1/2 } &= 0, \\
    \label{eq:LSA_BC_5_abbrev}
    \Theta_{f} \left( z= 1/2 \right) &= \Theta_{st} \left( z=1/2 \right), \\
    \label{eq:LSA_BC_6_abbrev}
    \Theta_{f} \left( z= -1/2 \right) &= \Theta_{sb} \left( z= -1/2 \right), \\
    \label{eq:LSA_BC_7_abbrev}
    D \Theta_{f} |_{z=1/2} &= \frac{\lambda_{st}}{\lambda_{f}} D \Theta_{st} |_{z=1/2}, \\
    \label{eq:LSA_BC_8_abbrev}
    D \Theta_{f} |_{z=-1/2} &= \frac{\lambda_{sb}}{\lambda_{f}} D \Theta_{sb} |_{z=-1/2}.
\end{align}

\subsection{Marginally stable state}
\label{sec:LSA_Marginally_stable_state}

The marginally stable state is the one we are looking for, defining neutral stability. Depending on the imaginary part of the growth rate, the system may be in overstability, if for at least one wavenumber $\Imag \left(  \sigma \right) \neq 0 $. Otherwise, if all wavenumbers result in an imaginary part of $0$, the principle of exchange of stabilities is valid.

\citet{Chandrasekhar1981} has shown that the latter applies for the Rayleigh-Bénard convection set-up. Moreover, Hurle has adapted this to the CHT case, allowing for the application of this principle \citep{Hurle1967}. Thus, we can set
\begin{align}
    \label{eq:LSA_sigma_zero}
    \Real \left(  \sigma \right) \stackrel{!}{=} 0 \land \Imag \left(  \sigma \right) \stackrel{!}{=} 0 \Rightarrow \sigma \equiv 0.
\end{align} 
Applying this to \eqref{eq:LSA_NSE_final} -- \eqref{eq:LSA_EE_s_final} gives us our modelling equations. We can re-write \eqref{eq:LSA_EE_f_final}
\begin{align}
    \label{eq:LSA_Theta_W_relation}
    \underbrace{\Pr \sigma \Theta_{f}}_{\sigma=0} = \left( D^2 -k^2 \right) \Theta_{f} + W \Leftrightarrow \Theta_{f} = - \frac{W}{D^2-k^2} \thickspace \thickspace \textrm{or} \thickspace \thickspace W = - \left( D^2-k^2 \right) \Theta_{f} ,
\end{align}
and insert it into \eqref{eq:LSA_NSE_final}, yielding either an ordinary differential equation (ODE) for $W$
\begin{align}
    \label{eq:LSA_W_modelling}
    \left( \left( D^2-k^2 \right)^3 + \Ra k^2\right) W = 0
\end{align}
or, alternatively, for $\Theta_{f}$
\begin{align}
    \label{eq:LSA_Theta_f_modelling}
    \left( D^2 -k^2 \right) \left( \left( D^2 -k^2 \right)^3 + \Ra k^2 \right) \Theta_{f} = 0 .
\end{align} 
For the solids, in turn, 
\begin{align}
    \label{eq:LSA_Theta_s_modelling}
    \underbrace{\Pr \kapparatioInv \sigma \Theta_{s}}_{\sigma=0} &=  \left( D^2 - k^2 \right) \Theta_{s} \Leftrightarrow \left( D^2 -k^2 \right) \Theta_{s} = 0.
\end{align}
An interesting aspect of equations \eqref{eq:LSA_W_modelling} -- \eqref{eq:LSA_Theta_s_modelling} is that the dependence on the Prandtl number $\Pr$ cancels out (because we assumed $\sigma \equiv 0$). Furthermore, all three equations are linear ordinary partial differential equations, making them comparatively easy to solve.


\subsection{Solution of the modelling equations}
\label{sec:LSA_Solution_of_the_modelling_equations}

Dealing with linear ODEs, an exponential ansatz of the form
\begin{align}
    \label{eq:LSA_exp_ansatz}
    \phi \left( z \right) = e^{qz}, \quad \phi = \{ W, \Theta_{f}, \Theta_{sb}, \Theta_{st} \}
\end{align}
with the free parameter $q$ is usually promising. 
Starting with the simplest of the three equations for the  temperature deviation, that in the solids $\Theta_{sb,st}$ \eqref{eq:LSA_Theta_s_modelling}, we get
\begin{align}
    \label{eq:LSA_Theta_s_general_sol}
    \left( D^2 -k^2 \right) \Theta_{s} = 0 \Rightarrow \Theta_{s} \left( z \right) = c_1 e^{-kz} + c_2 e^{kz}.
\end{align}
At the very top and bottom of the plates, respectively, temperature deviations must be $0$ as the temperature is set by virtue of the (external) thermal boundary conditions. Thus, we apply
\begin{alignat}{2}
    \label{eq:LSA_Theta_st_sol}
    &\Theta_{st} \left( z=1/2+\Gst \right)     &&\xRightarrow{\eqref{eq:LSA_Theta_s_general_sol}} \Theta_{st} \left( z \right) = c_1 \left( e^{-kz} - e^{kz} e^{-k(1+2\Gst)} \right), \\
    \label{eq:LSA_Theta_sb_sol}
    &\Theta_{sb} \left( z=- \left (1/2+\Gsb \right) \right)     &&\xRightarrow{\eqref{eq:LSA_Theta_s_general_sol}} \Theta_{sb} \left( z \right) = c_2 \left( e^{kz} - e^{-kz} e^{-k(1+2\Gsb)} \right).
\end{alignat}
Solving the equations \eqref{eq:LSA_W_modelling} and \eqref{eq:LSA_Theta_f_modelling} with this the exponential ansatz leads to
\begin{alignat}{2}
    \label{eq:LSA_W_ansatz}
    &\left( \left( q^2 - k^2 \right)^3 + \Ra \, k^2  \right) W &&= 0 , \\
    \label{eq:LSA_Theta_f_ansatz}
    \left( q^2 -k^2 \right) &\left( \left( q^2 -k^2 \right)^3 + \Ra\, k^2 \right) \Theta_{f} &&= 0 
\end{alignat}
whereas we obtain
\begin{align}
    \label{eq:LSA_Theta_f_W_relationship}
    \Theta_{f} = \frac{\left( q^2 -k^2 \right)^2}{\Ra k^2} W
\end{align}
using \eqref{eq:LSA_NSE_final} when setting $\sigma=0$ and using the same exponential ansatz. We obtain the solutions
\begin{align}
    \label{eq:LSA_q15}
    q_{1,5} &= \pm \sqrt{k^2 - \left( \Ra \, k^2 \right)^{\frac{1}{3}}}, \\
    \label{eq:LSA_q26}
    q_{2,6} &= \pm \sqrt{k^2 + 1/2 \left( \Ra \, k^2 \right)^{\frac{1}{3}}\left( 1-\mathrm{i}\sqrt{3} \right)}, \\
    \label{eq:LSA_q37}
    q_{3,7} &= \pm \sqrt{k^2 + 1/2 \left( \Ra \, k^2 \right)^{\frac{1}{3}}\left( 1+\mathrm{i}\sqrt{3} \right)}, \\
    \label{eq:LSA_q48}
    q_{4,8} &= \pm k,
\end{align}
where $q_{4,8}$ applies to \eqref{eq:LSA_Theta_f_ansatz} only whereas the other solutions are valid for both \eqref{eq:LSA_W_ansatz} and \eqref{eq:LSA_Theta_f_ansatz}.

Looking at the $\left( q^2 - k^2 \right)$-term in \eqref{eq:LSA_W_ansatz} and \eqref{eq:LSA_Theta_f_ansatz}, one can observe symmetry. Therefore, the solution may be written as a combination of $\sinh$ and $\cosh$ with orthogonal basis functions, splitting the overall solution into the even and odd solutions
\begin{align}
    \label{eq:LSA_W_ansatz_symmetric}
    W &= W_e + W_o ~~~~~ = \sum_{j=1}^3 A_j \cosh{\left( q_j z \right) } + \sum_{j=1}^3 B_j \sinh{\left( q_j z \right)}, \\
    \label{eq:LSA_Theta_F_ansatz_symmetric_with_Ta}
    \Theta_{f} &= \Theta_{f,e} + \Theta_{f,o} = \sum_{j=1}^4 C_j \cosh{\left( q_j z \right) } + \sum_{j=1}^4 D_j \sinh{\left( q_j z \right)} \\
    \label{eq:LSA_Theta_f_ansatz_symmetric}
    \Theta_{f} &= \frac{1}{\Ra \, k^2} \left( \sum_{j=1}^3 A_j \left( q_j^2-k^2 \right)^2 \cosh{\left( q_j z \right) } + \sum_{j=1}^3 B_j \left( q_j^2-k^2 \right)^2 \sinh{\left( q_j z \right)} \right),
\end{align}
where \eqref{eq:LSA_Theta_f_W_relationship} is used to obtain \eqref{eq:LSA_Theta_f_ansatz_symmetric}.


\subsection{Applying the boundary conditions}
\label{sec:LSA_Applying_the_boundary_conditions}

Now we apply the boundary conditions \eqref{eq:LSA_BC_12_abbrev} -- \eqref{eq:LSA_BC_8_abbrev} to solve \eqref{eq:LSA_W_ansatz_symmetric} and \eqref{eq:LSA_Theta_F_ansatz_symmetric_with_Ta} while using these conditions for the solids \eqref{eq:LSA_Theta_st_sol} and \eqref{eq:LSA_Theta_sb_sol}. For each condition, the equations \eqref{eq:LSA_W_ansatz_symmetric} and \eqref{eq:LSA_Theta_F_ansatz_symmetric_with_Ta} are split into even and odd modes and solved separately. Regarding identical bottom and top plates, we can further set
\begin{align}
    \label{eq:LSA:lambda_sb_is_lambda_st}
    \lambda_{st} = \lambda_{sb} = \lambda_{\mathrm{s}}, \\
    \Gst = \Gsb = \Gs.
\end{align}
Note that this makes the boundary conditions for the solid bottom domain obsolete as the solution coincides with that of the solid top domain by virtue of symmetry. For \eqref{eq:LSA_BC_12_abbrev}, one obtains 
\begin{align}
    \label{eq:LSA_BC1_sol_even}
    W_e \left(z=1/2 \right) &= \sum_{j=1}^3 E_j = E_1 + E_2 + E_3 = 0, \\
    \label{eq:LSA_BC1_sol_odd}
    W_o \left(z=1/2 \right) &= \sum_{j=1}^3 O_j = O_1 + O_2 + O_3 = 0.
\end{align}
with the new constants
\begin{align}
    \label{eq:LSA_E_j_definition}
    E_j&:=A_j \cosh{\left( \frac{q_j}{2} \right)}, \\
    \label{eq:LSA_O_j_definition}
    O_j&:=B_j \sinh{\left( \frac{q_j}{2} \right)}.
\end{align}
Applying \eqref{eq:LSA_BC_34_abbrev} yields
\begin{align}
    \label{eq:LSA_BC3_sol_even}
    DW_e \left( z=1/2 \right) &= \sum_{j=1}^3 E_j q_j t_j = 0, \\
    \label{eq:LSA_BC3_sol_odd}
    DW_o \left( z=1/2 \right) &= \sum_{j=1}^3 O_j q_j c_j = 0,
\end{align}
with
\begin{align}
    \label{eq:LSA_t_j_definition}
    t_j := \tanh{\left( \frac{q_j}{2} \right)}, \\
    \label{eq:LSA_c_j_definition}
    c_j := \coth{\left( \frac{q_j}{2} \right)} .
\end{align}
For \eqref{eq:LSA_BC_5_abbrev} one gets
\begin{align}
    \label{eq:LSA_BC5_sol_even}
    -E_0 \rev{\lambdaratioInv} \tanh{\left( k \rev{\Gs} \right)} + \frac{1}{\Ra \, k^2} \sum_{j=1}^3 E_j \left( q_j^2-k^2 \right)^2 &= 0, \\
    \label{eq:LSA_BC5_sol_odd}
    -O_0 \rev{\lambdaratioInv} \tanh{\left( k \rev{\Gs} \right)} + \frac{1}{\Ra \, k^2} \sum_{j=1}^3 O_j \left( q_j^2-k^2 \right)^2 &= 0.
\end{align}
where the new constant
\begin{align}
    \label{eq:LSA_E_0_definition}
    E_0 = O_0 := c_1 e^{-k/2} \cdot \rev{\lambdaratio} \cdot \frac{1-e^{-2k\rev{\Gs}}}{\tanh{\left( k \rev{\Gs} \right)}} .
\end{align}
Finally, for \eqref{eq:LSA_BC_7_abbrev} one obtains
\begin{align}
    \label{eq:LSA_BC7_sol_even}
    E_0k + \frac{1}{\Ra \, k^2}  \sum_{j=1}^3 E_j q_j t_j \left( q_j^2-k^2 \right)^2 &= 0, \\
    \label{eq:LSA_BC7_sol_odd}
    O_0k + \frac{1}{\Ra \, k^2} \sum_{j=1}^3 O_j q_j c_j \left( q_j^2-k^2 \right)^2 &= 0.
\end{align}

As a result, we are left with two systems of equations: One for the even modes and one for the odd modes. Together with the parameter
\begin{align}
    \label{eq:LSA_gamma_definition}
     \gamma := \left( \Ra k^2 \right)^{-1/3} ,
\end{align}
these two equations \eqref{eq:LSA_BC7_sol_even} and \eqref{eq:LSA_BC7_sol_odd} can be written as
\begin{alignat}{3}
    \label{eq:LSA_M_E_simplified}
    &\underbrace{
    \begin{pmatrix}
        0 & 1 & 1 & 1 \\
        0 & q_1t_1 & q_2t_2 & q_3t_3 \\
        - \lambdaratioInv \tanh{\left(k\Gs \right)} & \gamma & -\frac{\gamma}{2} \left( 1-\mathrm{i} \sqrt{3} \right)  & -\frac{\gamma}{2} \left( 1-\mathrm{i} \sqrt{3} \right) \\
        k & \gamma q_1t_1 & -\frac{\gamma}{2} \left( 1-\mathrm{i} \sqrt{3} \right) q_2t_2 & -\frac{\gamma}{2} \left( 1-\mathrm{i} \sqrt{3} \right) q_3t_3
    \end{pmatrix}
    }_{\bm{M_E}}
    &&\underbrace{
    \begin{pmatrix}
        E_0 \vspace{0.166cm} \\ E_1 \vspace{0.166cm} \\ E_2 \vspace{0.166cm} \\ E_3
    \end{pmatrix}
    }_{\bm{E}}
    &&=
    \begin{pmatrix}
        0 \vspace{0.166cm} \\ 0 \vspace{0.166cm} \\ 0 \vspace{0.166cm} \\ 0
    \end{pmatrix} \\
    \label{eq:LSA_M_O_simplified}
    &\underbrace{
    \begin{pmatrix}
        0 & 1 & 1 & 1 \\
        0 & q_1c_1 & q_2c_2 & q_3c_3 \\
        - \lambdaratioInv \tanh{\left(k\Gs \right)} & \gamma & -\frac{\gamma}{2} \left( 1-\mathrm{i} \sqrt{3} \right)  & -\frac{\gamma}{2} \left( 1-\mathrm{i} \sqrt{3} \right) \\
        k & \gamma q_1c_1 & -\frac{\gamma}{2} \left( 1-\mathrm{i} \sqrt{3} \right) q_2c_2 & -\frac{\gamma}{2} \left( 1-\mathrm{i} \sqrt{3} \right) q_3c_3
    \end{pmatrix}
    }_{\bm{M_O}}
    &&\underbrace{
    \begin{pmatrix}
        O_0 \vspace{0.166cm} \\ O_1 \vspace{0.166cm} \\ O_2 \vspace{0.166cm} \\ O_3
    \end{pmatrix}
    }_{\bm{O}}
    &&=
    \begin{pmatrix}
        0 \vspace{0.166cm} \\ 0 \vspace{0.166cm} \\ 0 \vspace{0.166cm} \\ 0
    \end{pmatrix}
\end{alignat} 
In other words, both the even and odd solutions represent a system of $4$ equations which must be solved at once. 
Note that the involved $q_j$, $t_j$, $c_j$, and $\gamma$ are defined in equations \eqref{eq:LSA_q15} -- \eqref{eq:LSA_q37}, \eqref{eq:LSA_t_j_definition}, \eqref{eq:LSA_c_j_definition}, and \eqref{eq:LSA_gamma_definition}, respectively.
The trivial solutions, i.e., all constants $E_j,O_j=0 \thickspace \forall \thickspace j \in [ 1,4 ]$, are obvious. To find the non-trivial solution, the determinants of the given coefficient matrices $\bm{M_E}, \bm{M_O}$ must contract to zero. 

In this solution, there are $4$ free variables left: $\Gs$, $\lambdaratioInvIL$, $k$, and $\Ra$. This allows to fix $\Gs$ and $\lambdaratioInvIL$ to compute the neutral stability curves $\Ra \left( k \right)$ for the even and odd solutions. 

We find that odd solutions are generally more stable than even solutions. In other words, the neutral stability curves from even solution are below those from odd solutions. Hence, we restrict our analysis in the main text, see equation \eqref{eq:det_ME}, to the solution of equation \eqref{eq:LSA_M_E_simplified}.


\section{Convergence of thermal boundary conditions at extreme ratios of thermal diffusivities $\kapparatioIL$}
\label{sec:Appendix_convergence}

\begin{figure}
\centering
\includegraphics[width = \textwidth]{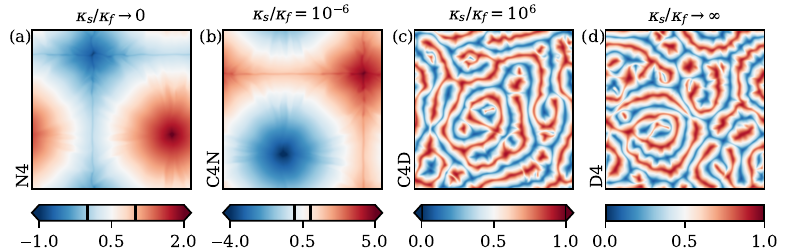}
\caption{\justifying{
Pattern formation at extreme ratios of thermal diffusivities $\kapparatioIL$. 
For extreme values of $\kapparatioIL$, (b, c) the flow patterns of CHT simulations converge perfectly towards (a, d) those obtained from classical idealised thermal boundary conditions.
Here $\Ra = 10^{4}$ and $\Gs = 15$ (if applicable). Note that the simulations from panels (a, d) do not comprise any solid plates, whereas those from panels (b, c) do.
}}
\label{fig:CHT_validation}
\end{figure}

In order to validate our CHT set-up -- which adds two identical fluid-confining solid plates at the top and bottom of the classically studied fluid layer --, we conduct two simulations using extreme $\kapparatioIL = \left\{ 10^{-6}, 10^{6} \right\}$. These parameters are supposed to resemble the idealised Neumann and Dirichlet case, $\kapparatioIL \rightarrow 0$ and $\kapparatioIL \rightarrow \infty$, respectively. 

For understanding these limits and their consequences, it is helpful to think of our CHT set-up as a thermal circuit with its different involved thermal resistances or conductances (representing, e.g., the different sub-domains) \citep{Incopera2007, Vieweg2025}. If one pathway of heat transfer offers a significantly smaller thermal resistance $R_{th}$ -- or, as $R_{th} \sim \lambda^{-1} \sim \kappa^{-1}$, larger thermal conductivity or diffusivity --, it represents a shortcut and will be favoured over other pathways.

In case of $\kapparatioIL \rightarrow 0$ or $\kappa_{\textrm{s}} \ll \kappa_{\textrm{f}}$, the thermal resistance of the solid plates is enormous. However, as our domain is horizontally periodic, heat transfer through these plates is unavoidable and the \enquote{easiest} way through the plate is by passing it vertically. 
This means for the present set-up that heat transfer from $\Th$ to $\Tc$ tends to avoid \enquote{expensive} horizontal transfer within the solid plates and results in a uniform heat flux at the solid-fluid interfaces. Thus, the classical Neumann case is resembled.

In the opposite case of $\kapparatioIL \rightarrow \infty$ or $\kappa_{\textrm{s}} \gg \kappa_{\textrm{f}}$, the thermal resistance of the solid plates is tiny whereas that of the fluid layer is huge. Heat transfer starting from $\Th$ or towards $\Tc$ will go all possible ways within the solid plates before eventually interacting with the resistive fluid layer. As a result, heat transfer within the solid plates is faster than through the fluid layer and the plates become iso-thermal, leading to the classical Dirichlet case.

Figure \ref{fig:CHT_validation} compares our extreme CHT cases with classical, plate-less scenarios. It is obvious that the more natural implementations of the Neumann and Dirichlet cases match their idealised counterparts very well. This first qualitative impression is further confirmed by $\Nu$, $\Re$, and $\intLS$ as provided in table \ref{tab:simulation_outcome_characteristics}. 

The direct comparison of a plate-less Neumann case and the plate-involving CHT case results in an artefact concerning the mean temperature difference across the fluid layer:
In the Neumann case, the non-dimensionalisation is not based on the characteristic temperature $T_{char,D} = \dT$ (as in the Dirichlet and CHT cases) but rather on the applied vertical temperature gradient across the fluid layer $T_{char,N} = - \partial T / \partial z$. In other words, the characteristic temperature scale is different. As a result, the resulting mean temperature difference $\dTN \leq 1$ \citep{Otero2002}. 
This leads to out-of-line scaling in the legend of figure \ref{fig:CHT_validation} as well as $\maxDhT$ and $\stdT$ in table \ref{tab:simulation_outcome_characteristics}. 
However, this is more of a technical issue and can be circumvented by re-scaling the solution fields based on $\dTN$ \citep{Vieweg2023a, Vieweg2024a}, eventually re-aligning the values and proving the validity of our used set-up.

\bibliographystyle{jfm}

\end{document}